\documentclass[conference, 10pt]{IEEEtran}
\IEEEoverridecommandlockouts
\usepackage{cite}
\usepackage{amsmath,amssymb,amsfonts}
\usepackage{hyperref}
\usepackage{listings}
\usepackage{algorithmic}
\usepackage{graphicx}
\usepackage[normalem]{ulem}
\usepackage{textcomp}
\usepackage{physics}
\usepackage{float} 
\usepackage{caption}
\usepackage{placeins}
\usepackage{subcaption}
\usepackage{tikz}
\usetikzlibrary{positioning} %
\usepackage{adjustbox}
\usetikzlibrary{3d,calc, shapes.geometric, arrows.meta}
\usepackage{tikz-3dplot}
\def\BibTeX{{\rm B\kern-.05em{\sc i\kern-.025em b}\kern-.08em
    T\kern-.1667em\lower.7ex\hbox{E}\kern-.125emX}}
\usepackage{xcolor}
\definecolor{codegray}{rgb}{0.5,0.5,0.5}
\definecolor{backcolour}{rgb}{0.97,0.97,0.97}

\newcommand{\linebreakand}{%
  \end{@IEEEauthorhalign}
  \hfill\mbox{}\par
  \mbox{}\hfill\begin{@IEEEauthorhalign}
}

\lstdefinestyle{mystyle}{
    backgroundcolor=\color{white},        
    commentstyle=\color{gray}\itshape,    
    keywordstyle=\color{blue},            
    stringstyle=\color{orange},           
    basicstyle=\ttfamily\footnotesize,    
    mathescape=true,  
    breakatwhitespace=false,         
    breaklines=true,                 
    captionpos=b,                    
    keepspaces=true,                 
    numbers=none,                    
    showspaces=false,                
    showstringspaces=false,
    showtabs=false,                  
    frame=single,
    rulecolor=\color{black},
    tabsize=2
}

\begin{document}

\title{MPS-JuliQAOA: User-friendly, Scalable MPS-based Simulation for Quantum Optimization \\
}

\newcommand{\rt}[1]{{\color{blue}[(Reuben) #1}]}
\newcommand{\rte}[1]{{\color{blue} #1}}
\newcommand{\cut}{\text{cut}}

\definecolor{kellygreen}{rgb}{0.3, 0.73, 0.09}

\newcommand{\se}[1]{{\color{kellygreen}[(Stephan) #1}]}
\newcommand{\see}[1]{{\color{kellygreen} #1}}

\newcommand{\sfe}[1]{{\color{red}[(Sean) #1}]}
\newcommand{\sfee}[1]{{\color{red} #1}}

\author{
\IEEEauthorblockN{Sean Feeney}
\IEEEauthorblockA{\textit{CCS-3 Information Sciences} \\
\textit{Los Alamos National Laboratory}\\
Los Alamos, NM, 87544 USA \\
sfeeney@lanl.gov}
\and
\IEEEauthorblockN{Reuben Tate}
\IEEEauthorblockA{\textit{CCS-3 Information Sciences} \\
\textit{Los Alamos National Laboratory}\\
Los Alamos, NM, 87544 USA \\
rtate@lanl.gov}
\\[1.5ex]
\IEEEauthorblockN{Stephan Eidenbenz}
\IEEEauthorblockA{\textit{CCS-3 Information Sciences} \\
\textit{Los Alamos National Laboratory}\\
Los Alamos, NM, 87544 USA \\
eidenben@lanl.gov}
\and
\IEEEauthorblockN{John Golden}
\IEEEauthorblockA{\textit{CCS-3 Information Sciences} \\
\textit{Los Alamos National Laboratory}\\
Los Alamos, NM, 87544 USA \\
golden@lanl.gov}

}

\maketitle

\begin{abstract}
We present the MPS-JuliQAOA simulator, a user-friendly, open-source tool to simulate the Quantum Approximate Optimization Algorithm (QAOA) of any optimization problem that can be expressed as diagonal Hamiltonian. By leveraging Julia-language constructs and the ITensor package to implement a Matrix Product State (MPS) approach to simulating QAOA, MPS-Juli-QAOA effortlessly scales to 512 qubits and 20 simulation rounds on the standard de-facto benchmark 3-regular MaxCut QAOA problem. MPS-JuliQAOA also has built-in parameter finding capabilities, which is a crucial performance aspect of QAOA. We illustrate through examples that the user does not need to know MPS principles or complex automatic differentiation techniques to use MPS-JuliQAOA. We study the scalability of our tool with respect to runtime, memory usage and accuracy tradeoffs. Code available at \href{https://github.com/lanl/JuliQAOA.jl/tree/mps}{https://github.com/lanl/JuliQAOA.jl/tree/mps}.
\end{abstract}

\begin{IEEEkeywords}
MPS, QAOA, Max-Cut
\end{IEEEkeywords}

\section{Introduction}

Quantum computing is a promising candidate method to overcome current scaling and accuracy challenges in the fields of  computation, physics, chemistry, and optimization \cite{di2024quantum,cao2019quantum}. While large-scale fully Fault-Tolerant Quantum Computers (FTQCs) are expected to be available by 2030 (see roadmaps by most vendors, e.g., IBM's roadmap https://www.ibm.com/roadmaps/quantum/), we can now use classical HPC simulations of quantum circuits to understand 
 what problems can be solved on a quantum computer, and study anticipated behavior \cite{xu2023herculeantaskclassicalsimulation}.

Though the discovery of quantum algorithms is an ongoing research effort, many prime candidates have already been proposed. 
Algorithms such as Shor's algorithm for prime number factorization \cite{365700} provide an exponential speedup over their classical counterparts; others, such as Grover's unstructured search algorithm \cite{10.1145/237814.237866} provide a quadratic speedup. Variational quantum algorithms, which parameterize a quantum circuit and use a classical optimizer outer-loop to search for optimal parameters that maximize or minimize some cost function and an inner quantum loop to evolve the quantum circuit, have become promising candidates for research into a variety of problems in optimization, machine learning, ground-state finding, and many other problems \cite{Peruzzo_2014, farhi2014quantumapproximateoptimizationalgorithm, cerezo2021variational}. While known worst-case performance bounds are typically not the strength of variational algorithms 
they are likely to be used as practical heuristics that  outperform classical algorithms on real-world problem instances, akin to the continued use of classical heuristics (such as simulated annealing or the simplex algorithm) over theoretically more robust alternatives. 

The exact approach to (classically) simulating these methods is state vector simulation. State vector simulation provides an exact solution to what a quantum circuit would output on a Fault Tolerant Quantum Computer (FTQC). To some extent, the runtime of state vector simulation can be mitigated via parallel processing and GPU utilization,
however, runtime limits are rarely the bottleneck for larger systems. Quantum bits (qubits) exist in a $2^{n}$ dimensional space, meaning that, in order to faithfully represent the state vector, classical computers have exponentially-growing memory requirements as the qubit count grows; moreover, the approaches described above are not sufficient for simulating large systems in the regime of hundreds of qubits.
%
To overcome memory limitation issues, a few clever algorithms have been designed, most of which involve the implementation of Tensor Network (TN) architectures \cite{berezutskii2025tensor}. Popular architectures for such formulations include the 1-D Matrix Product State (MPS), the generalized MPS Projected Entangled Pairs States (PEPS) for 2D systems and higher \cite{verstraete2004renormalization}, and Multiscale Entanglement Renormalization Ansatz (MERA) \cite{cincio2008multiscale}.
Many algorithms have been put forward for efficient simulation of these architectures. Such families of algorithms include the Density Matrix Renormalization Group (DMRG), which is a variational approach to find lowest-energy matrix product state wavefunction \cite{PhysRevLett.69.2863}. Other time evolution methods include Time-evolving block decimation (TEBD) and Time-dependent variational principle (TDVP)  \cite{PAECKEL2019167998}. Algorithms like DMRG, TEBD, and TDVP offer a promising toolkit for researchers looking to simulate lowly-entangled systems of hundreds, or even thousands of qubits. The speed of these algorithms is due to the nature of their designs: tensor contractions are innately structured for parallel processing on both CPUs and GPUs. Yet, there is no free lunch: these algorithms, while robust in there construction, fast in the speed, and memory friendly, are (typically) themselves approximations of the quantum state. Parameters such as singular value cutoff and bond-dimension ($\chi$) allow researchers to determine the approximation and memory tradeoffs necessary to simulate quantum systems.

In this paper we present a Matrix Product State (MPS) simulation toolkit that accepts as input, a general problem Hamiltonian, and mixer for the Quantum Approximation Optimization Algorithm (QAOA) \cite{farhi2014quantumapproximateoptimizationalgorithm}; a variational quantum algorithm for approximately solving discrete optimization problems \cite{farhi2014quantumapproximateoptimizationalgorithm}. Building on Golden et. al \cite{Golden_2023} \texttt{JuliQAOA}, we present in this paper \texttt{MPS-JuliQAOA}. We show that this tool drastically scales up the number of qubits from 10s to 100s. We also show several unique features that make the tool easy to use and implement on Nvidia GPUs as well as multi-threading and automatic-differentiation capabilities. We first present a background to the reader in Section \ref{sec:Background} on the basics of Tensor Networks and MPS, before introducing the standard MaxCut optimization problem as well as the QAOA method, and Automatic-Differentiation that is used in the outer loop of QAOA. We describe the methods in Section \ref{sec:software_implemenation}, giving on overview of the architecture and workflow. Building on this workflow we show code examples in Section \ref{sec:SoftwareUsage}, displaying the simple to use interface and abstraction for users. Finally we present the runtime and memory scaling results as well as GPU and Multi-threading capabilities in Section \ref{sec:Results}. Concluding remarks are in Section \ref{sec:conclusion}.

\section{Related Work}
A variety of QAOA simulators have been introduced in the last several years. Implementations vary from the well established Qiskit tool kit \cite{qiskit2024} to \texttt{QAOA.jl} implemented by Bode et al. \cite{Bode2023} and based on their previous algorithmic work \cite{PRXQuantum.4.030335}. Other implementations such as Golden et al. \cite{Golden_2023} and the \texttt{JuliQAOA.jl} software provide state-of-the-art state vector simulation capability and include advanced automatic-differentiation methods for parameter optimization. However, due the fundamental nature of the exponential memory scaling requirements of quantum systems, these software packages fall short in providing the scalability necessary for researchers to simulate large-scale quantum systems greater than $40$ qubits.

As a result of state vector simulation memory limitations, tensor networks, and their associated architectures such as MPS and PEPS, have grown in popularity in recent years. New software libraries such as \texttt{ITensor} \cite{ITensor} and \texttt{QTensor} \cite{Lykov_2021} have been developed to fill these needs. However, these libraries often have a very steep learning curve to use effectively, i.e. users must be familiar with tensor networks, their structure, and the associated algorithms like TEBD or DMRG, on top of an understanding of QAOA and its circuit implementation, as well as advanced parameter optimization methods such as automatic-differentiation. 

Due to these challenges, there is a high barrier-to-entry for researchers who wish to study optimization problems using QAOA with an MPS framework. As a result, there is a lack of a unified software implementation for users to simply specify their cost Hamiltonian and run QAOA MPS simulations without having to implement it from scratch themselves. Our proposed MPS-JuliQAOA tool fills that gap. MPS-JuliQAOA retains the modeling philosophy developed by Golden et al. \cite{Golden_2023} in \texttt{JuliQAOA.jl}, ut significantly expands the scalability beyond CPU state vector simulation.

\section{Background}
\label{sec:Background}
\subsection{Tensor Networks}
Tensor networks are structured representations of high-dimensional tensors that use a graphical notation to simplify the description and manipulation of complex systems \cite{Bridgeman_2017}. In this notation, tensors are depicted as shapes with legs representing indices, and connecting legs indicate contraction, or summation, over shared indices. This approach generalizes Einstein summation notation and provides a visual, intuitive way to manage the exponential complexity of high dimensional systems.

\begin{figure}[h]
    \centering
    \begin{tikzpicture}[scale=1.2, every node/.style={scale=1}]
        \node[draw, minimum width=1.2cm, minimum height=1cm] (A) at (0,0) {$A$};
        \draw[-] (A.west) -- ++(-1,0) node[left] {$\alpha$};
        \draw[-] (A.east) -- ++(1,0) node[right, above] {$\beta$};

        \node[draw, minimum width=1.2cm, minimum height=1cm] (B) at (3,0) {$B$};
        \draw[-] (B.west) -- ++(-1,0);
        \draw[-] (B.east) -- ++(1,0) node[right] {$\gamma$};

        \draw[-] (A.east) -- (2,0) -- (B.west);

        \node[draw, minimum width=1.2cm, minimum height=1cm] (C) at (1.5,-2) {$C$};
        \draw[-] (C.west) -- ++(-1,0) node[left] {$\alpha$};
        \draw[-] (C.east) -- ++(1,0) node[right] {$\gamma$};

        \draw[dashed, ->] (A.south) -- (C.north west);
        \draw[dashed, ->] (B.south) -- (C.north east);
    \end{tikzpicture}
    \caption{Matrix multiplication as a tensor contraction. The shared index \( \beta \) is contracted, resulting in a new tensor with open indices \( \alpha \) and \( \gamma \).}
    \label{fig:tensor-contraction}
\end{figure}
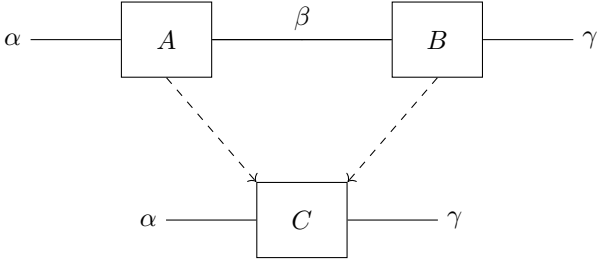

A simple example of this is matrix multiplication, which can be viewed as a tensor contraction \cite{biamonte2017tensornetworksnutshell}. Two rank-2 tensors, $A_{\alpha\beta}$ and $B_{\beta\gamma}$, are connected along their shared index $\beta$, producing a new tensor $C_{\alpha\gamma}$ (Equation \ref{eqn:tensor_example}). In tensor diagrammatic notation, this is represented by linking the corresponding legs of $A$ and $B$, and the resulting shape in Figure \ref{fig:tensor-contraction} retains the uncontracted indices. This illustrates how standard linear algebra operations naturally fit into the tensor network framework, and can be highly useful in the representation of quantum systems.

\begin{equation}
    C_{\alpha\gamma} = \sum_{\beta} A_{\alpha\beta} B_{\beta\gamma}
    \label{eqn:tensor_example}
\end{equation}

In quantum computing, tensor networks are particularly valuable for representing and simulating many-body quantum states\cite{Or_s_2014}. They allow for efficient storage and computation by capturing the entanglement structure of quantum systems without needing to represent the full Hilbert space explicitly. This is especially useful for systems obeying area-law entanglement, where the relevant quantum information can be compressed into a network of low-rank tensors.

\subsection{Matrix Product States}

Matrix Product States (MPS) are a widely used class of tensor networks particularly suited for representing quantum states of one-dimensional systems. An MPS expresses a global many-body quantum state as a sequence of contracted local tensors, each associated with a physical site (i.e., a qubit) and internal “bond” indices. This decomposition is especially effective for quantum systems that obey area-law entanglement, where the full state vector grows exponentially in size, but its entanglement structure allows for an efficient low-rank representation \cite{Verstraete_2006}.

Consider a pure quantum state \( |\psi\rangle \in (\mathbb{C}^2)^{\otimes n} \), which can be written in the computational basis as:
\begin{equation}
|\psi\rangle = \sum_{i_1, i_2, \dots, i_n = 0}^1 c_{i_1 i_2 \dots i_n} |i_1 i_2 \dots i_n\rangle.
\end{equation}
In the MPS representation, the coefficients \( c_{i_1 i_2 \dots i_n} \) are decomposed into a product of tensors:
\begin{equation}
c_{i_1 i_2 \dots i_n} = \sum_{\alpha_1, \dots, \alpha_{n-1}} A^{[1]}_{i_1 \alpha_1} A^{[2]}_{\alpha_1 i_2 \alpha_2} \cdots A^{[n]}_{\alpha_{n-1} i_n}.
\end{equation}
Each tensor \( A^{[k]} \) has one physical index \( i_k \) (associated with the qubit state) and up to two bond indices \( \alpha_{k-1}, \alpha_k \) that are contracted with neighboring tensors. For open boundary conditions, the first and last tensors have only one bond index.

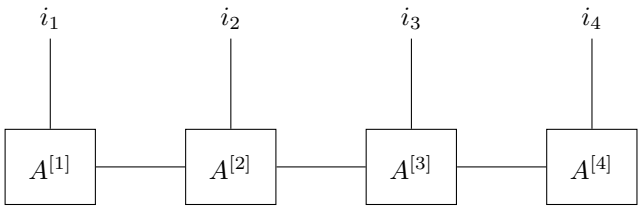
\begin{figure}[h]
    \centering
    \begin{tikzpicture}[scale=1.2, every node/.style={scale=1}]
        \foreach \i in {1,2,3,4} {
            \node[draw, minimum width=1.2cm, minimum height=1cm] (A\i) at (2*\i, 0) {$A^{[\i]}$};
            \draw[-] (A\i.north) -- ++(0,1) node[above] {$i_{\i}$};
        }

        \foreach \i/\j in {1/2,2/3,3/4} {
            \draw[-] (A\i.east) -- (A\j.west);
        }
    \end{tikzpicture}
    \caption{Matrix Product State (MPS) representation of an \( n=4 \) qubit quantum state. Each tensor \( A^{[k]} \) has one physical index \( i_k \) and is connected by bond indices to neighboring tensors.}
    \label{fig:mps-diagram}
\end{figure}

This structure makes MPS a natural and efficient format for simulating quantum circuits and time evolution in one-dimensional systems. Moreover, as a subclass of tensor networks, MPS inherits the graphical notation introduced earlier. In this language, an MPS is drawn as a chain of tensors with horizontal legs (bonds) connecting adjacent tensors and vertical legs (physical indices) representing local degrees of freedom (i.e. 2-D for qubits), as shown in Figure~\ref{fig:mps-diagram}.

A key aspect of Matrix Product State (MPS) representations is the user's ability to tune the hyperparameters: the cutoff value and the bond dimension $\chi$. The MPS is constructed through the iterative application of Singular Value Decomposition (SVD) along the chain of tensors.

By adjusting the cutoff, users set a threshold below which singular values are discarded. This effectively removes small contributions to the entanglement between qubits, simplifying the representation while retaining the dominant correlations in the system.

The bond dimension $\chi$ places a hard limit on the number of singular values retained on the virtual, or “bond,” indices between adjacent tensors. While users can control either the cutoff or $\chi$, a common practice is to set a small cutoff value (e.g., $10^{-12}$) and then incrementally increase $\chi$. This allows users to balance computational cost with the fidelity of the representation, retaining only the most significant entanglement features without exceeding manageable tensor sizes. Often, $\chi$ is increased in powers of 2 to systematically explore this trade-off.

\subsection{Classical Max-Cut}

The Max-Cut problem is a fundamental combinatorial optimization problem. Given a graph \( G = (V, E) \) with edge weights \( w: E \rightarrow \mathbb{R} \), the goal is to partition the vertex set \( V \) into two disjoint subsets such that the sum of the weights of the edges crossing the partition is maximized. For a graph with \( |V| = n \), a feasible solution can be encoded as a bitstring \( b \in \{0, 1\}^n \), where the value of the \( j \)-th bit indicates which subset vertex \( j \) belongs to.

Formally, the Max-Cut problem can be expressed as:
\begin{equation}
\text{Max-Cut}(G) = \max_{b \in \{0,1\}^n} \text{cut}(b),
\end{equation}
where
\begin{equation}
\label{eqn:maxcut}
\text{cut}(b) = \frac{1}{2} \sum_{(i,j) \in E} w_{ij} \cdot \mathbf{1}[b_i \neq b_j].
\end{equation}
Here, \( \mathbf{1}[b_i \neq b_j] \) is the indicator function that evaluates to 1 if vertices \( i \) and \( j \) are assigned to different subsets, and 0 otherwise.

\subsection{Quantum Approximate Optimization Algorithm (QAOA)}

The Quantum Approximate Optimization Algorithm (QAOA) is a variational quantum algorithm designed to approximate solutions to combinatorial optimization problems, such as Max-Cut. Given a classical objective function \( f : \{0,1\}^n \rightarrow \mathbb{R} \), QAOA seeks to prepare a quantum state \( |\psi\rangle \) from which high-quality solutions can be sampled.

QAOA is defined by the following components:
\begin{itemize}
    \item A cost Hamiltonian \( H_C \) that is diagonal in the computational basis and encodes the classical objective function: \( H_C |x\rangle = f(x) |x\rangle \),
    \item A mixing Hamiltonian \( H_M \),
    \item A positive integer \( p \) representing the number of alternating layers (or depth),
    \item Two parameter vectors \( \boldsymbol{\gamma} = (\gamma_1, \ldots, \gamma_p) \) and \( \boldsymbol{\beta} = (\beta_1, \ldots, \beta_p) \),
    \item An initial state \( |s_0\rangle \), typically a uniform superposition over all bitstrings.
\end{itemize}

The QAOA ansatz is constructed by alternating applications of the cost and mixing unitaries:
\begin{equation}
|\boldsymbol{\gamma}, \boldsymbol{\beta} \rangle = e^{-i\beta_p H_M} e^{-i\gamma_p H_C} \cdots e^{-i\beta_1 H_M} e^{-i\gamma_1 H_C} |s_0\rangle.
\end{equation}
Each cost unitary applies a phase based on the classical objective, while each mixing unitary drives transitions between computational basis states. A classical optimizer is then used to tune \( \boldsymbol{\gamma} \) and \( \boldsymbol{\beta} \) in order to maximize the expectation value \( \langle \boldsymbol{\gamma}, \boldsymbol{\beta} | H_C | \boldsymbol{\gamma}, \boldsymbol{\beta} \rangle \), guiding the quantum state toward high-quality solutions.

\subsection{Automatic Differentiation}
A key aspect of QAOA is the classical outer-loop, in which the parameters $\gamma$ and $\beta$ are optimized to maximize some metric, usually expectation value, but can also be implementation or problem specific such as in \cite{feeney2024bettersolutionprobabilitymetric}. There are many methods that can be used to optimize $\gamma$ and $\beta$. Gradient-free optimizers are often used when gradients are unavailable or are too computationally expensive to calculate; however they require many more iterations to converge when compared to gradient-based algorithms, and may not converge at all \cite{kus2024gradientfreeneuraltopologyoptimization}. In situations where the gradient can be efficiently computed, numerical differentiation methods using finite differences can be a powerful tool. Nevertheless, numerical differentiation methods suffer from round-off and truncation errors \cite{baydin2018automaticdifferentiationmachinelearning}. They are also known to struggle with the computation of partial derivatives of functions with many inputs, which is necessary for high depth QAOA circuits.

Automatic Differentiation (AD), also known as algorithmic differentiation, is a method for computing the exact gradients of a function $f(x)$. It is essentially the generalized form of the backpropagation algorithm, now ubiquitous in machine learning and neural network implementations across industry and academia. AD, in its simplest definition, breaks down complex functions into elementary operations or components forming a computational graph, and then through the iterative application of the chain rule (from calculus), is able to compute the exact gradients of the sub components of $f(x)$ and thus $f(x)$ itself. These gradients are then fed into a local optimizer algorithm like L-BFGS or Stochastic Gradient Descent, and the process is continued until convergence or some threshold value is met.  While the exact implementation characteristics of AD vary and can be highly involved, many surveys and reviews on these details can be found throughout the literature \cite{baydin2018automaticdifferentiationmachinelearning} \cite{fang2024stepbystepintroductionimplementationautomatic}. Relevant to this work is the implementation of AD methods for MPS and variational quantum algorithms \cite{Guo_2023}.

\section{Methods}
\label{sec:software_implemenation}
In this section we describe an overview of the implementation details for MPS representation and the construction of the QAOA circuit. We outline the acceptance of generalized diagonal cost Hamiltonians for optimization problems and relate it to the parameter optimization loop necessary for QAOA circuit performance. 
\subsection{Implementation Overview}

To implement the Quantum Approximate Optimization Algorithm (QAOA) for Max-Cut and generalized cost Hamiltonians, we use the Julia programming language due to its strong performance characteristics and support for differentiable programming. Our simulation framework is built on the \texttt{ITensors.jl} and \texttt{ITensorMPS.jl} libraries, which provide efficient tensor network data structures and operations, including Matrix Product State (MPS) representations.

The QAOA circuit is simulated using an MPS-based approach, where the initial state \( |s_0\rangle \) is represented as a product state and evolved under a sequence of unitary gates derived from the cost Hamiltonian \( H_C \) and mixing Hamiltonian \( H_M \). The cost Hamiltonians considered include the Max-Cut Hamiltonian as well as other diagonal cost functions defined over bitstrings \( x \in \{0, 1\}^n \).
\subsection{General Hamiltonian}
\label{sec:methods_hamiltonian}

Beyond the Max-Cut problem, our implementation supports any $n$-qubit Hamiltonian that can be expressed as a sum of Pauli-$Z$ products, i.e., any Hamiltonian of the form:
\begin{align}
\label{eqn:ham_form}
H_C &=  \sum_{\substack{S \subseteq [n]: \\ S \neq \varnothing }} \left(k_S \cdot \prod_{s \in S} Z_s\right)\\
&= \sum_{i} k_i Z_i + \sum_{i < j} k_{ij} Z_iZ_j + \sum_{i < j < k} k_{ijk} Z_iZ_jZ_k + \cdots \notag
\end{align}
where the $k_S$'s, $k_i$'s, $k_{ij}$'s, and $k_{ijk}$'s are real-valued coefficients, and the ``$+\cdots$" indicates potential higher-order terms.

For many Hamiltonians of interest, amongst all possible subsets $S \subseteq [n] = \{1,2,\dots, n\}$, most of the $k_S$ coefficients will be 0; our software only requires the user to specify the terms with non-zero coefficients. If the Hamiltonian is $d$-local, meaning that $k_S = 0$ for all $|S| > d$ (and there exists $S$ with $|S| = d$ such that $k_S \neq 0$), then the user will need to specify up to ${n \choose d} = O(n^d)$ terms. In particular, Max-Cut can be expressed with a 2-local Hamiltonian, and thus requires up to $O(n^2)$ terms.


The simulation framework accepts user-defined Hamiltonians structured as dictionaries or operator lists, making it extensible to a wide variety of combinatorial optimization problems and spin models.

\subsection{Parameter Optimization with Automatic Differentiation}

To optimize the QAOA parameters \( \boldsymbol{\gamma}, \boldsymbol{\beta} \), we use reverse mode automatic-differentiation via the \texttt{Zygote.jl} library. This enables gradient-based optimization of the expectation value \( \langle \boldsymbol{\gamma}, \boldsymbol{\beta} | H_C | \boldsymbol{\gamma}, \boldsymbol{\beta} \rangle \) with respect to the circuit parameters. Because the ITensor library supports custom adjoints and differentiable tensor operations, we are able to efficiently backpropagate through the full tensor network contraction representing the QAOA state.

Our method proceeds as follows:
\begin{enumerate}
    \item Initialize the circuit parameters \( \boldsymbol{\gamma}, \boldsymbol{\beta} \) and construct the initial product state \( |s_0\rangle \).
    \item Apply \( p \) alternating layers of phase and mixing unitaries using tensor network operators.
    \item Compute the expectation value of the cost Hamiltonian \( H_C \) on the resulting state.
    \item Use \texttt{Zygote.jl} to compute gradients of the expectation value with respect to \( \boldsymbol{\gamma}, \boldsymbol{\beta} \).
    \item Update parameters using a local or global optimizer(e.g., Basinhopping, L-BFGS).
    \item Repeat until convergence or a fixed number of steps is reached.
\end{enumerate}

\subsection{Software Architecture}

\begin{figure}[h]
    \centering
    \begin{tikzpicture}[
        node distance=1.5cm and 2.3cm,
        every node/.style={draw, rectangle, fill=gray!10, align=center, font=\small, text width=3.0cm},
        ]
        \node (input) {Define Cost Hamiltonian\\(e.g., Max-Cut)};
        \node (init) [left=of input] {Initialize MPS State \\ $|s_0\rangle$};
        \node (circuit) [below=of init] {Apply QAOA Layers \\ $e^{-i\beta H_M}, e^{-i\gamma H_C}$};
        \node (expectation) [below=of circuit] {Compute Cost Expectation \\ $\langle H_C \rangle$};
        \node (gradient) [right=of expectation] {Compute Gradients \\ (Zygote.jl)};
        \node (optimize) [right=of circuit] {Update Parameters \\ (Classical Optimizer)};

        \draw[->] (input) -- (init);
        \draw[->] (init) -- (circuit);
        \draw[->] (circuit) -- (expectation);
        \draw[->] (expectation) -- (gradient);
        \draw[->] (gradient) -- (optimize);
        \draw[->] (optimize.west) --  (circuit.east);

    \end{tikzpicture}
    \caption{Software flow for QAOA implementation using Julia, ITensor, and Zygote.}
    \label{fig:software-flow}
\end{figure}
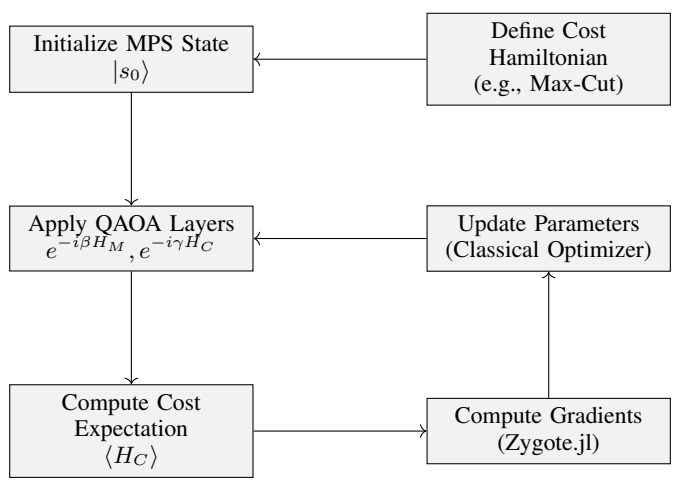

The structure of our QAOA implementation is modular, with clear separation between tensor network representation, circuit construction, and optimization. Figure~\ref{fig:software-flow} provides an overview of the software workflow.

\subsection{GPU and Multi threading Capabilities}
To accelerate MPS QAOA simulations, our implementation leverages both CPU multi-threading and GPU computation, dependent upon user preference. On the CPU side, we exploit two levels of parallelism: BLAS multithreading, which accelerates dense linear algebra routines through optimized multi-core CPU libraries, and ITensor's threaded block-sparse operations, which enable parallel tensor contractions within block-sparse tensor networks. While these two forms of parallelism can independently offer significant performance gains, they should not be used simultaneously without care, as excessive thread contention may degrade performance. In our benchmarks, BLAS multithreading generally provided superior performance for moderate bond dimensions and circuit depths, whereas threaded block-sparse operations proved beneficial in specific contraction-heavy workloads.

In addition, we implement GPU acceleration using the \texttt{CUDA.jl} backend for ITensors. The cost and mixer gates, as well as MPS tensors and MPO Hamiltonians, are transferred to the GPU and executed with cuBLAS and cuTENSOR routines, allowing efficient simulation of larger systems within the memory constraints of modern GPUs. The framework supports hybrid workflows, enabling flexible switching between CPU and GPU backends based on available resources. We note that currently only single-node GPU capabilities are implemented and device-specific information such as the GPU model is logged for each experiment.

Together, these capabilities enable scalable MPS QAOA simulations across a range of architectures, from desktop CPUs to high-performance clusters with GPU nodes, all based on user preference.

\section{Software Usage}
In this section we provide code examples for users to interface with \texttt{MPS-JuliQAOA}. We show how users specify a custom Hamiltonian of Pauli-Z operators using the custom ZInteractions structure, and then further into a QAOA problem structure that defines the initial MPS state. We also define a few custom problem constructors for tha MaxCut and Maximum Independent Set problems. We then show how users post-process their results as the cost Hamiltonians have to be altered for only Z interactions. Lastly we present an example of angle finding in which the user merely sends in the QAOA problem, which returns the expectation value along with the optimal angle parameters $\gamma$ and $\beta$.
\label{sec:SoftwareUsage}
\subsection{Hamiltonian Specification}
As discussed in Section \ref{sec:methods_hamiltonian}, most optimization problems of interest can be expressed as a sum of products of Pauli-$Z$ matrices. Our software accepts any Hamiltonian that can be expressed in such a way.

In what follows, we provide examples on how to specify the Hamiltonian for two different optimization problems: weighted Max-Cut and Maximum Independent Set. Our library also includes helper functions for Max-SAT. We refer the reader to \cite{hadfield2021representation} and \cite{lucas2014ising} regarding additional guidance on how to express a variety of optimization problems in this sum-of-products form.

\subsubsection{Max-Cut Example}

As an example, consider the weighted Max-Cut problem. Letting $z_i$ be a $\{-1, +1\}$ decision variable that determines which side of the cut the $i$th vertex is on, Equation \ref{eqn:maxcut} can be rewritten as:
$$\cut(z) = \sum_{(i,j) \in E} \frac{1}{2} w_{ij}(1 - z_iz_j).$$

The corresponding cost Hamiltonian is obtained by simply ``promoting" the decision variables to Pauli-$Z$ matrices:
$$H_C = \sum_{(i,j) \in E} \frac{1}{2} w_{ij}(I - Z_iZ_j).$$

In the literature, it is common for researchers to use a ``simplified" Hamiltonian for Max-Cut:
$$H_C' = \sum_{(i,j) \in E} w_{ij} Z_iZ_j.$$

Let us consider a specific graph $G = (V,E)$ with three vertices, edge set $E = \{(1,2), (2,3)\}$, and with weights $w_{1,2} = 1$ and $w_{2,3} = 2$. The simplified cost Hamiltonian becomes:
$$H_C' = Z_1Z_2 + 2Z_2Z_3.$$

This Hamiltonian is encoded an array of \texttt{ZInteractions} objects in the code, and can be constructed in the software as seen below. Each of the two terms above correspond to a \texttt{ZInteractions} object. Note that if a \texttt{ZInteractions} does not have its coefficient specified, then it is assumed to be 1.\\
\begin{center}
\begin{minipage}{240pt}
\begin{lstlisting}[language=Python,numbers=left, caption={Constructing cost Hamiltonian for Max-Cut}, label={lst:basic}]
using JuliQAOA

interactions = [
    ZInteractions([1,2])
    ZInteractions([2,3], 2)
]
\end{lstlisting}
\end{minipage}
\end{center}

Our software also includes helper functions for certain optimization problems such as Max-Cut, which take as input the problem instance in some natural representation (e.g. a Julia Graph for Max-Cut) and returns the appropriate array of interactions. Below is a code-snippet that returns the same array of \texttt{ZInteractions} objects as above.

\begin{center}
\begin{minipage}{240pt}
\begin{lstlisting}[language=Python,numbers=left, caption={Constructing cost Hamiltonian for Max-Cut with helper function}, label={lst:basic}]
using JuliQAOA
using Graphs, SimpleWeightedGraphs

#Construct the graph
g = SimpleWeightedGraph(3)
add_edge!(g, 1, 2, 1.0)
add_edge!(g, 2, 3, 2.0)

#Obtain simplified Max-Cut Hamiltonian
#as an interactions list
interactions = maxcut_graph_to_zinteractions(g, weighted=true)
\end{lstlisting}
\end{minipage}
\end{center}

\subsubsection{Maximum Independent Set}
Given a graph $G = (V,E)$, the goal of the Maximum Independent Set (MIS) problem is to find the largest subset $S \subseteq V$ of vertices that form an independent set, i.e., there are no edges between any of the vertices in $S$. Let $x_i$ denote the $\{0,1\}$ decision variable that is 1 if and only if vertex $i$ is included in $S$. Constraints aside, the objective is the maximize the number of vertices, i.e., $\sum_{i \in V} x_i$.

We encode the constraints through the use of a penalty function; for each edge amongst vertices in $S$, we apply a penalty of $\lambda$. Observe that for any edge $e = (i,j) \in E$, that $x_ix_j = 1$ if and only if $e$ is an edge connecting vertices in $S$, and thus the overall penalty term for the whole graph can be written as $-\lambda \sum_{(i,j) \in E} x_i x_j$. By picking $\lambda > 1$, any gains obtained from including constraint-violating vertices is more than offset by the penalty incurred.

Let $z_i$ be the corresponding $\{-1,+1\}$ decision variable that is obtained from the mapping $z_i = 1 - 2x_i$, or equivalently, $x_i = \frac{1}{2}(1- z_i)$. We add the objective and penalty terms, apply this mapping, and similar to Max-Cut, we then promote the $z_i$ variables to Pauli matrices; the result is as follows:
\begin{equation}\label{eqn:MIS_Ham} H_C = \frac{1}{2}\sum_{i \in V} (I - Z_i) - \frac{\lambda}{4} \sum_{(i,j) \in E} (I - Z_i)(I - Z_j).\end{equation}

This Hamiltonian requires some manipulation in order to put it in a form suitable for our software (see Equation \ref{eqn:ham_form}). Performing such manipulations yields the following:
$$H_C = kI + \sum_{i \in V} k_i Z_i + \sum_{(i,j) \in E} k_{ij} Z_iZ_j,$$
where,
$$k = \frac{n}{2} - \frac{\lambda}{4}m$$
$$k_i = \frac{\lambda  }{4}\text{deg}_i - \frac{1}{2}$$
$$k_{ij} = -\frac{\lambda}{4},$$
where $\text{deg}_i$ is the degree of vertex $i$ in the graph $G$. The derivation of these coefficient values can be found in Appendix \ref{sec:MIS_appendix}.

Our software includes a Hamiltonian converter that takes the MIS graph and calculates the above coefficients automatically; see the code snippet below which shows an example with a 3-node graph with a single edge. 

\begin{center}
\begin{minipage}{240pt}
\begin{lstlisting}[language=Python,numbers=left, caption={Constructing cost Hamiltonian for Max-Cut with helper function}, label={lst:basic}]
using JuliQAOA
using Graphs

#Construct the graph
g = SimpleGraph(3)
add_edge!(g, 1, 2)

#Obtain simplified Max-Cut Hamiltonian
#as an interactions list
interactions = mis_graph_to_zinteractions(g)
\end{lstlisting}
\end{minipage}
\end{center}

The converter ignores the identity term, and instead runs QAOA with the simplified Hamiltonian $H_C' = H_C - kI$. Just like with Max-Cut, we can also account for the removed identity term by postprocessing on the expectation value (which simply adds $k$ to the expectation), see Section \ref{sec:postprocessing}.

\subsection{Running MPS QAOA}
Once the interactions are specified, it is straightforward to then run MPS QAOA with specific variational parameters. The code snippet below illustrates how to do this with $p=2$ QAOA layers. The output is the MPS-approximated expectation value of the QAOA at those parameters. While the \texttt{run\_qaoa\_mps} function has key word arguments for the specification of both cutoff and bond dimension, users can override these default values by specifying the keywords with their own values.

\begin{center}
\begin{minipage}{240pt}
\begin{lstlisting}[language=Python,numbers=left, caption={Running MPS QAOA with specified variational parameters}, label={lst:basic}]
#Specify angles [betas..., gammas...]
beta_1 = 1.3
beta_2 = 2.4
gamma_1 = 0.7
gamma_2 = 3.1
angles = [beta_1, beta_2, gamma_1, gamma_2]

#Run MPS QAOA at specified angles
problem = QAOAProblem(interactions)
e_val = run_qaoa_mps(angles, problem; 
        cutoff=1e-10, maxdim=128)
\end{lstlisting}
\end{minipage}
\end{center}

\subsection{Post-Processing}
\label{sec:postprocessing}
If MPS-QAOA is ran with a simplified Hamiltonian $H_C'$ that is not equal to original Hamiltonian of the problem; the expectation value may need to be post-processed appropriately. Here, we illustrate how this is done with Max-Cut; the case for MIS and other optimization problems are similar.

The relationship between the original and simplified Hamiltonian for Max-Cut is as follows:
$$H_C = \frac{W}{2}I - \frac{1}{2}H_C',$$
where $W = \sum_{(i,j) \in E} w_{ij}$ is the sum of the weights of the edges. 

Thus, if QAOA is ran using $H_C'$ with parameters $(\gamma',\beta')$ and obtains an expectation of $\mathcal{E}'$, then QAOA ran using $H_C$ with parameters $(-2\gamma', \beta')$ will give an expectation of $$\mathcal{E} = \frac{W}{2} - \frac{1}{2}\mathcal{E}'.$$

For each problem-specific Hamiltonian converter, our software contains an associated helper function that allows the user to apply such post-processing on the expectation value as seen below.

\begin{center}
\begin{minipage}{240pt}
\begin{lstlisting}[language=Python,numbers=left, caption={Post-processing the Max-Cut expectation value}, label={lst:basic}]
#Post process to account for prefactors 
#in MaxCut Hamiltonian i.e. |W|/2 - e_val/2
e_val = maxcut_post_process(e_val; interactions=interactions)
\end{lstlisting}
\end{minipage}
\end{center}

\subsection{Angle Finding}
A major part of QAOA as a variational quantum algorithm is the classical outerloop. While the previous examples have shown how to perform MPS QAOA simualtion with given angles, it is often the case that users will want to find the best angles which maximize or minimize the cost Hamiltonian. Our implementation does this through Julia's rich eco-system of Automatic-Differentiation libraries. The user specifies the cost Hamiltonian, either through custom interactions, or through one of the problem specific interaction builders.  The user can choose to "warm start" the angle finding function with their own angles, if not the default angles are initialized randomly.

Once the interaction list is specified the user simply calls the angle finding function and a dictionary with the maximized or minimized energy, optimal angles, and several other metrics is returned. The number of rounds is specified by either p or the length of the angles divided by 2.
\label{sec:anglefindingexample}
\begin{center}
\begin{minipage}{240pt}
\begin{lstlisting}[language=Python,numbers=left, caption={Calculate optimal angles for specified QAOA problem}, label={lst:basic}]
#calculate optimal angles for MaxCut
#in MaxCut Hamiltonian i.e. |W|/2 - e_val/2
problem = QAOAProblem(interactions)

#intitializing empty dictionary for return
results = Dict() 

#"warm start" angles
angles=[1.57,3.14,3.14,1.57]
results = angle_finding_mps(problem; p=2, 
max=true, angles=angles)

e_val = results[:energy]
opt_angles = results[:angles]
\end{lstlisting}
\end{minipage}
\end{center}

\section{Results}
\label{sec:Results}
In the following section we present the scaling results for the MPS-JuliQAOA simulator described in Section \ref{sec:software_implemenation}. we present runtime and memory scaling results of the \texttt{MPS-JuliQAOA} software. We detail the implications of bond dimension $\chi$ and its impact on runtime, memory, and QAOA circuit depth. Finally we give a brief overview of the GPU and Multi-Threading scaling capabilities.
\subsection{Experimental setup}
\label{sec:experiment_setup}
 As a standard benchmark, we tested the simulator on the Max-Cut problem on 3-regular graphs. For each problem size $n \in \{32, 64, 128, 256\}$ we generate a fixed random 3-regular graph, and then simulate QAOA using MPS for varying bond dimensions ranging from  $ \chi \in \{ 64, 128, 256, 512\}$ and circuit depth $p \in \{1,2,3, \dots, 20\}$. For GPU and Multi-Threading we generate random 3-regular graphs of sizes $n \in 10,20,..,100$ and $p \in 1,2,3,4$.

\begin{figure*}[htp]
  \centering
  \subfloat[]{%
    \includegraphics[width=0.49\textwidth]{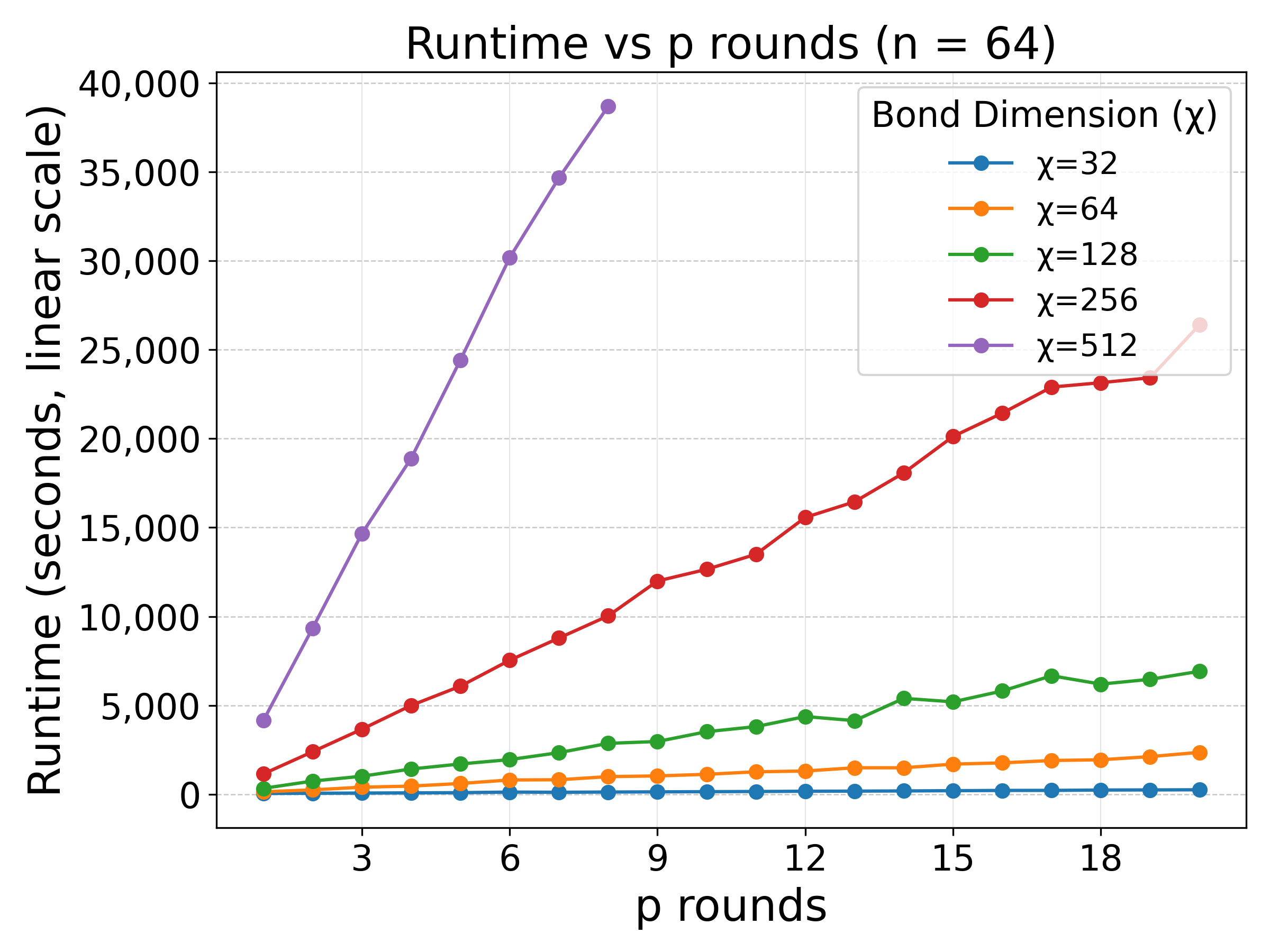} 
  }%
  \hfill
  \subfloat[]{%
    \includegraphics[width=0.49\textwidth]{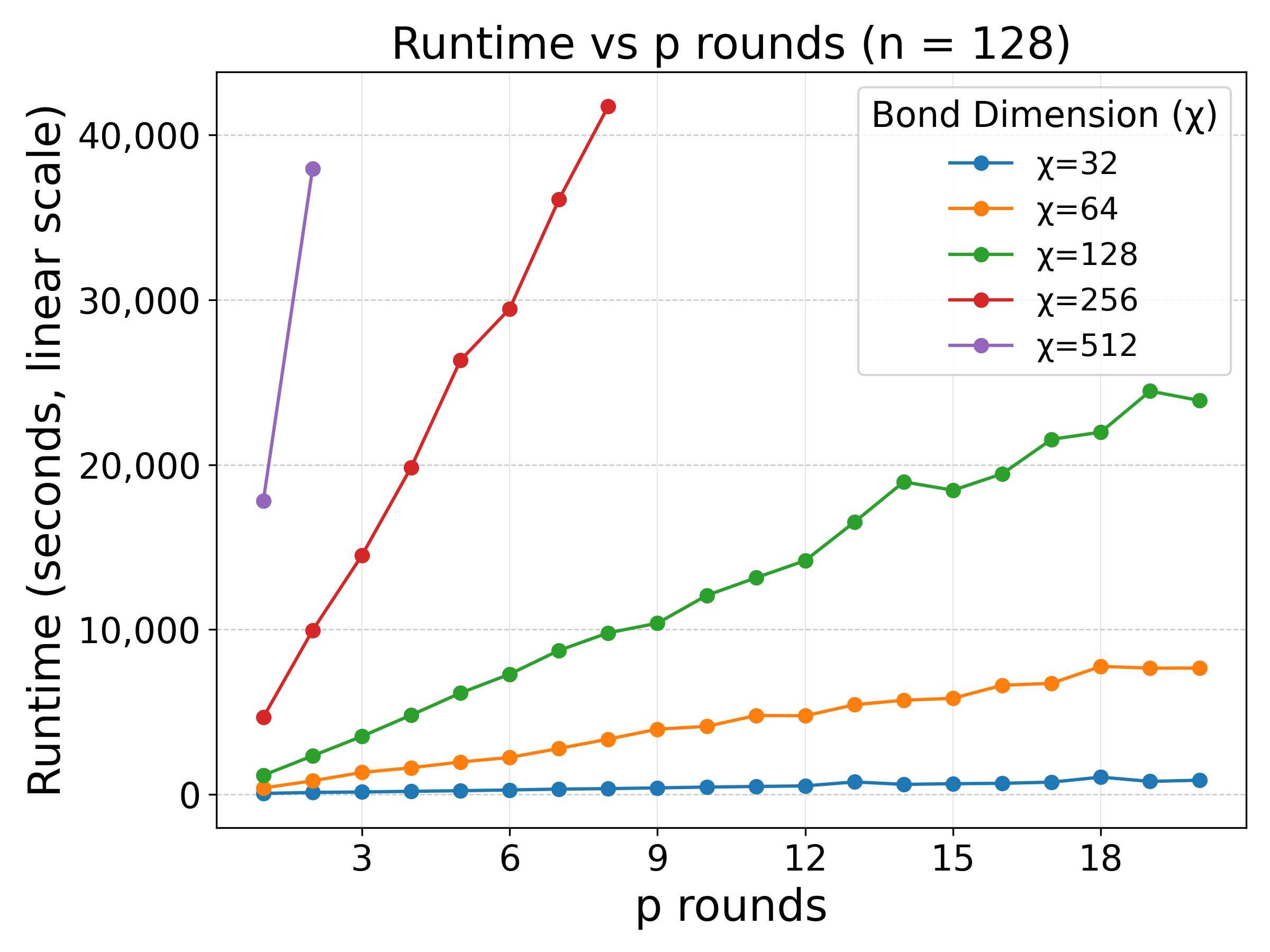}%
  }\\
  \vspace{0.cm} 
  \subfloat[]{%
    \includegraphics[width=0.49\textwidth]{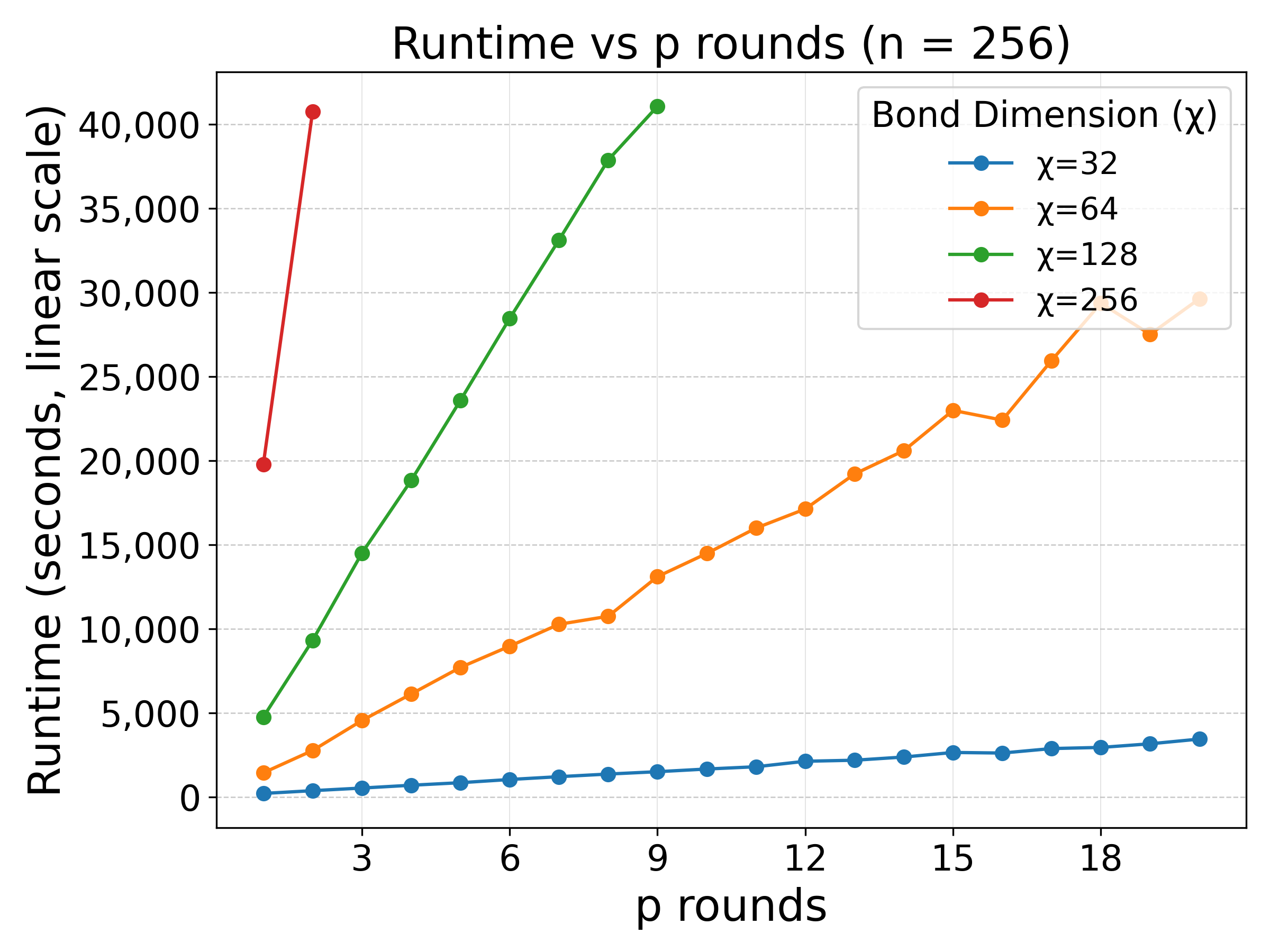}%
  }%
  \hfill
  \subfloat[ ]{%
    \includegraphics[width=0.49\textwidth]{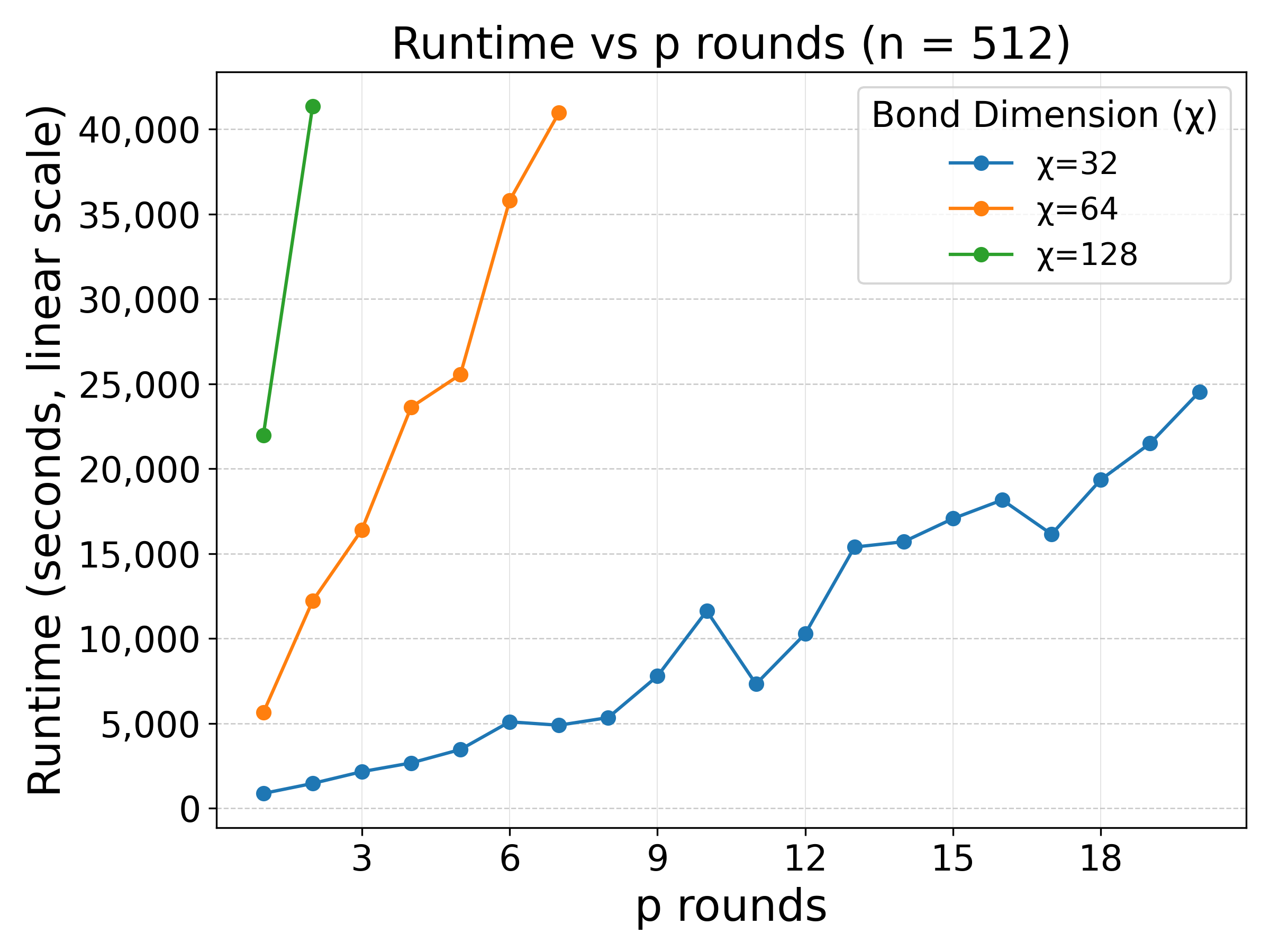}%
  }
  
  \caption{Circuit depth $p$ (x-axis) vs runtime (y-axis) for problem sizes $n\in\{ 64, 128, 256, 512\}$. Each curve represents a bond dimension $\chi \in \{32,64,128,256,512\}$.}
  \label{fig:runtime_vs_p}
\end{figure*}

\subsection{Runtime vs Circuit Depth}
We first evaluate the runtime performance of our QAOA MPS simulator across varying circuit depths and bond dimensions. Figure~\ref{fig:runtime_vs_p} depicts the runtime in seconds (y-axis) against the number of QAOA rounds $p$ (i.e., the circuit depth). Each subplot corresponds to a fixed problem size $n$, while each curve within a subplot shows performance for a particular bond dimension $\chi$.

Across all subplots in Figure \ref{fig:runtime_vs_p}, a consistent trend is observed: runtime increases with circuit depth $p$, and this growth is more pronounced at higher bond dimensions $\chi$. However, it can be observed that the relationship between circuit depth and runtime for a fixed value of $\chi$ appears to be linear. 

For example, in Figure~\ref{fig:runtime_vs_p}(a), corresponding to $n = 64$, we observe that at $\chi = 256$, the runtime increases from approximately 1000 seconds at $p = 1$ to nearly 10,000 seconds at $p = 8$, and continues to grow up to around 26,000 seconds by $p = 20$. This pattern of linear scaling holds across the other problem sizes as well.

\begin{figure*}[htp]
  \centering
  \subfloat[]{%
    \includegraphics[width=0.49\textwidth]{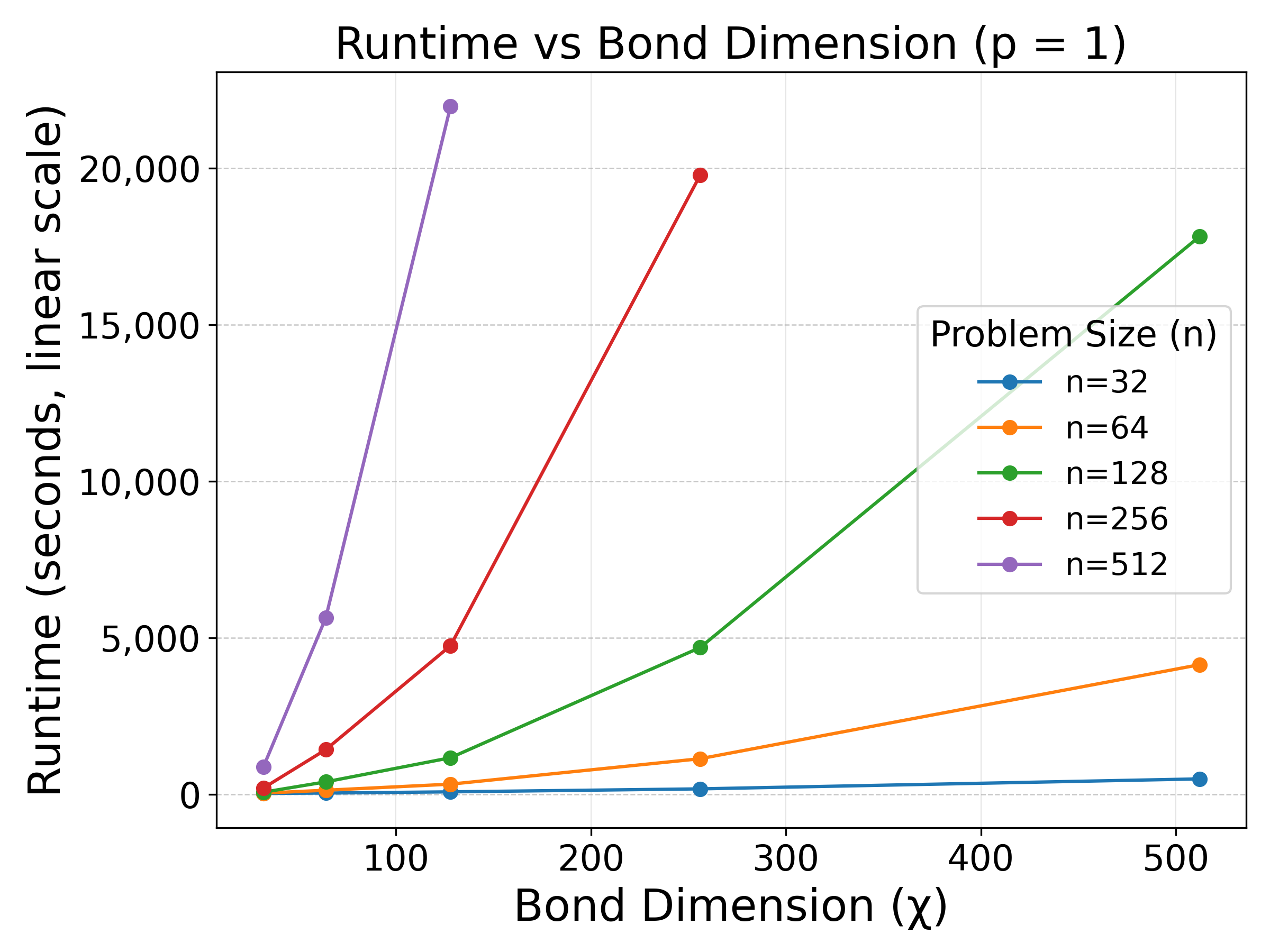} 
  }%
  \hfill
  \subfloat[]{%
    \includegraphics[width=0.49\textwidth]{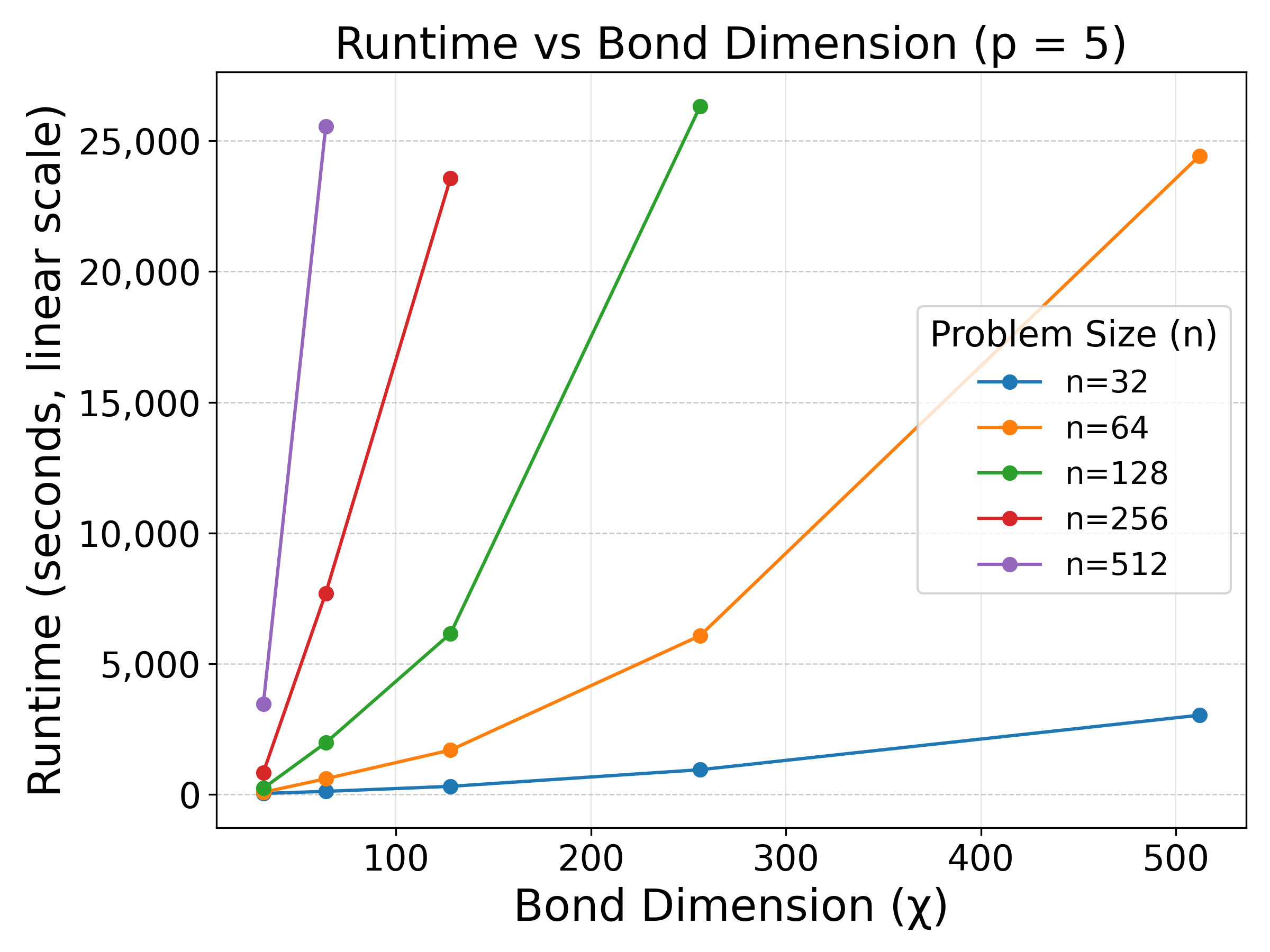}%
  }\\
  \vspace{0.cm} 
  \subfloat[]{%
    \includegraphics[width=0.49\textwidth]{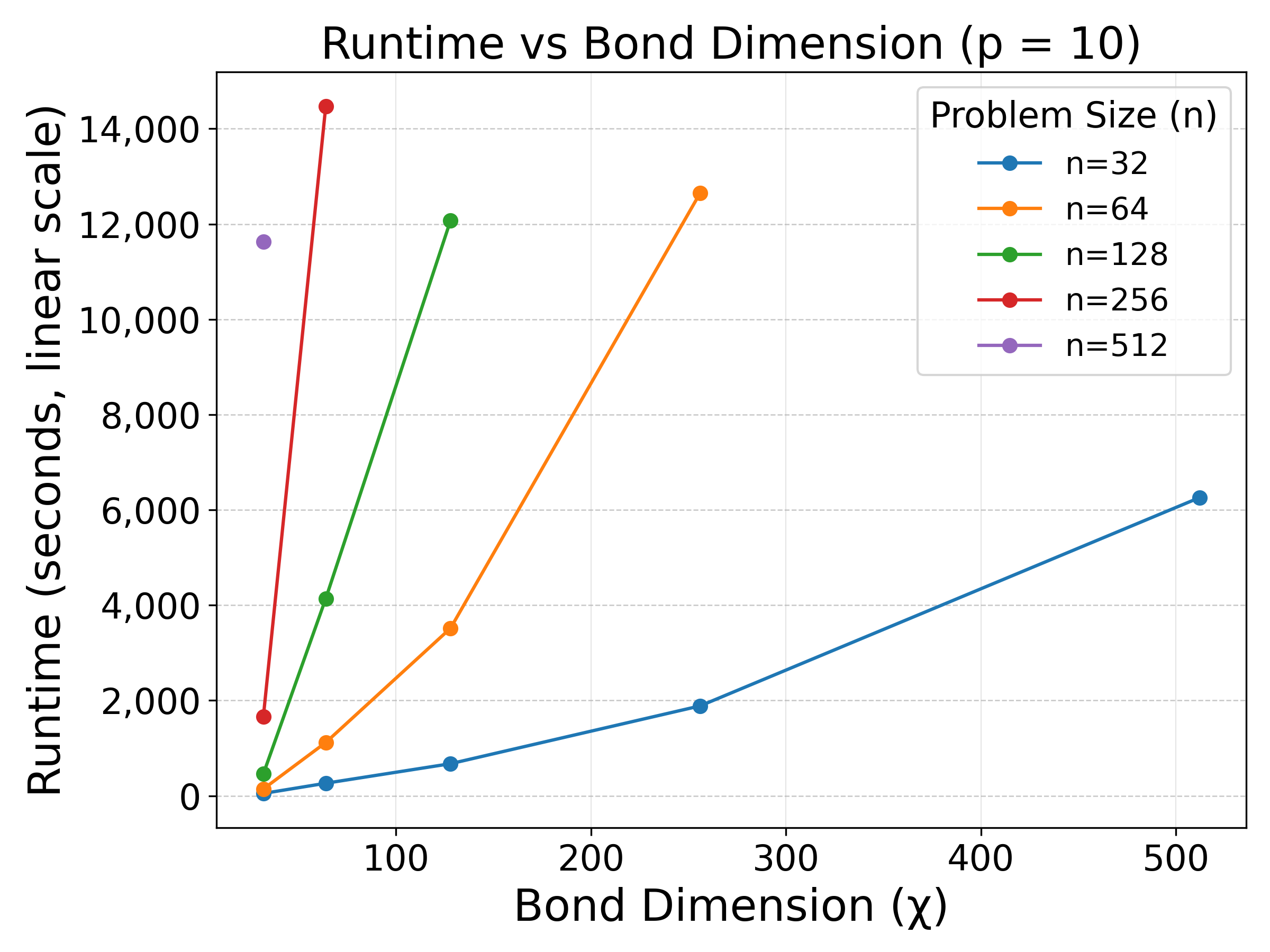}%
  }%
  \hfill
  \subfloat[ ]{%
    \includegraphics[width=0.49\textwidth]{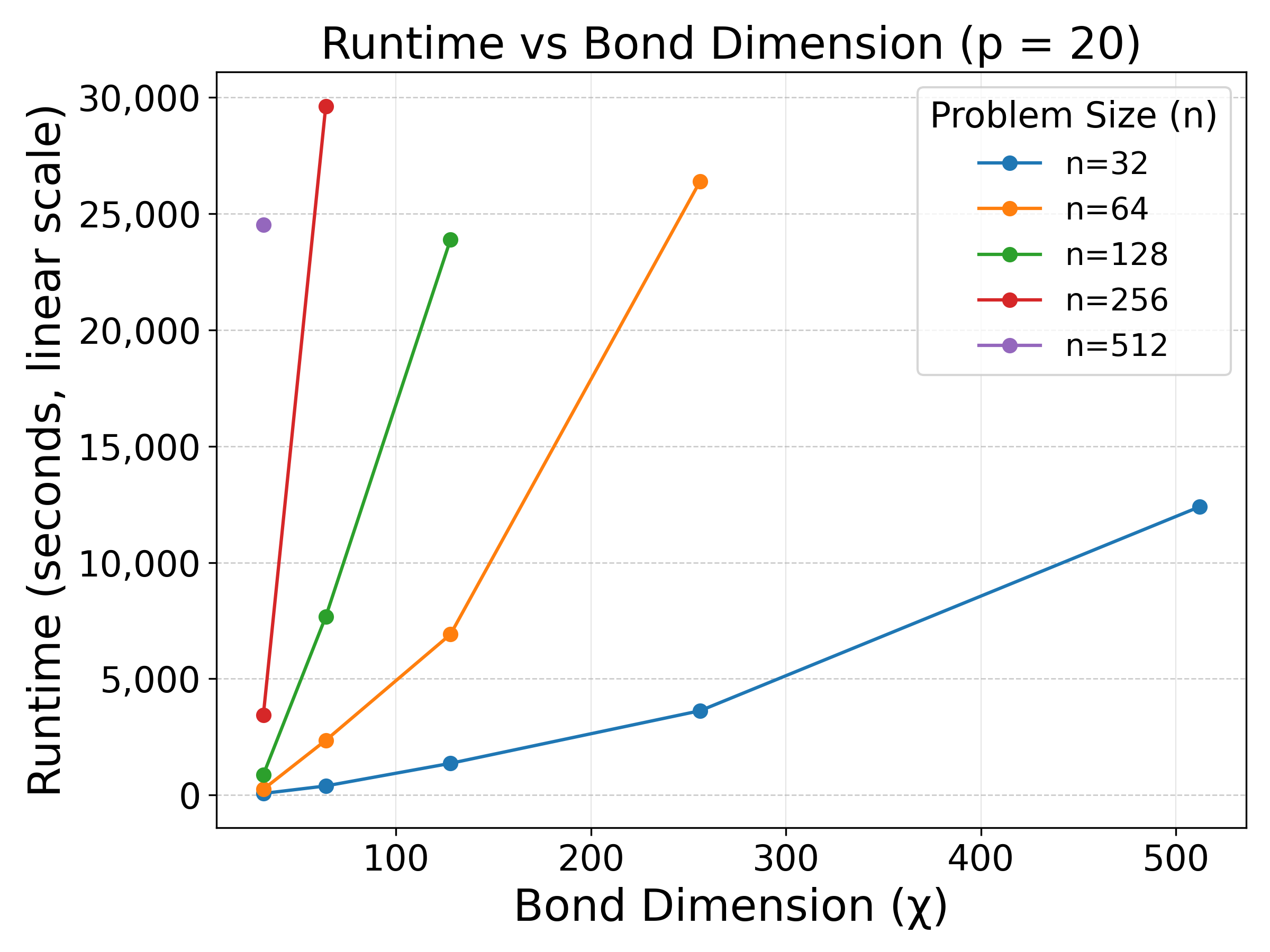}%
  }
  
  \caption{Bond dimension $\chi$ (x-axis) vs runtime (y-axis) for circuit depth $p\in \{1, 5, 10, 20\}$. Each curve represents a problem size $n \in \{32,64,128,256,512\}$. }
  \label{fig:bonddim_vs_runtime}
\end{figure*}

As we move from subplots (a) to (d) in Figure \ref{fig:runtime_vs_p}, increasing the number of qubits $n$, we also see a general upward shift in runtime across all $p$ and $\chi$ values. This is expected since simulating larger systems requires storing and manipulating more tensors and the absolute runtime increase with $n$ due to the larger number of qubits and hence tensor contractions, which are known to be the primary bottleneck with most tensor network approaches \cite{berezutskii2025tensor}. We also note the impact of $n$ on runtime is less steep than the impact of $\chi$.

Figure \ref{fig:runtime_vs_p} highlights the scalability of MPS-JuliQAOA with respect to both problem size $n$ and QAOA circuit depth $p$, but also reveals that the dominant factor influencing runtime is the bond dimension $\chi$, which governs the entanglement capacity of the MPS ansatz. Although increasing $n$ and $p$ naturally impacts runtime, it is the tuning of $\chi$ that presents a more nuanced challenge, balancing simulation fidelity against hardware and performance constraints. This critical trade-off is explored in detail in the following analysis.

\begin{figure*}[htp]
  \centering
  \subfloat[]{%
    \includegraphics[width=0.49\textwidth]{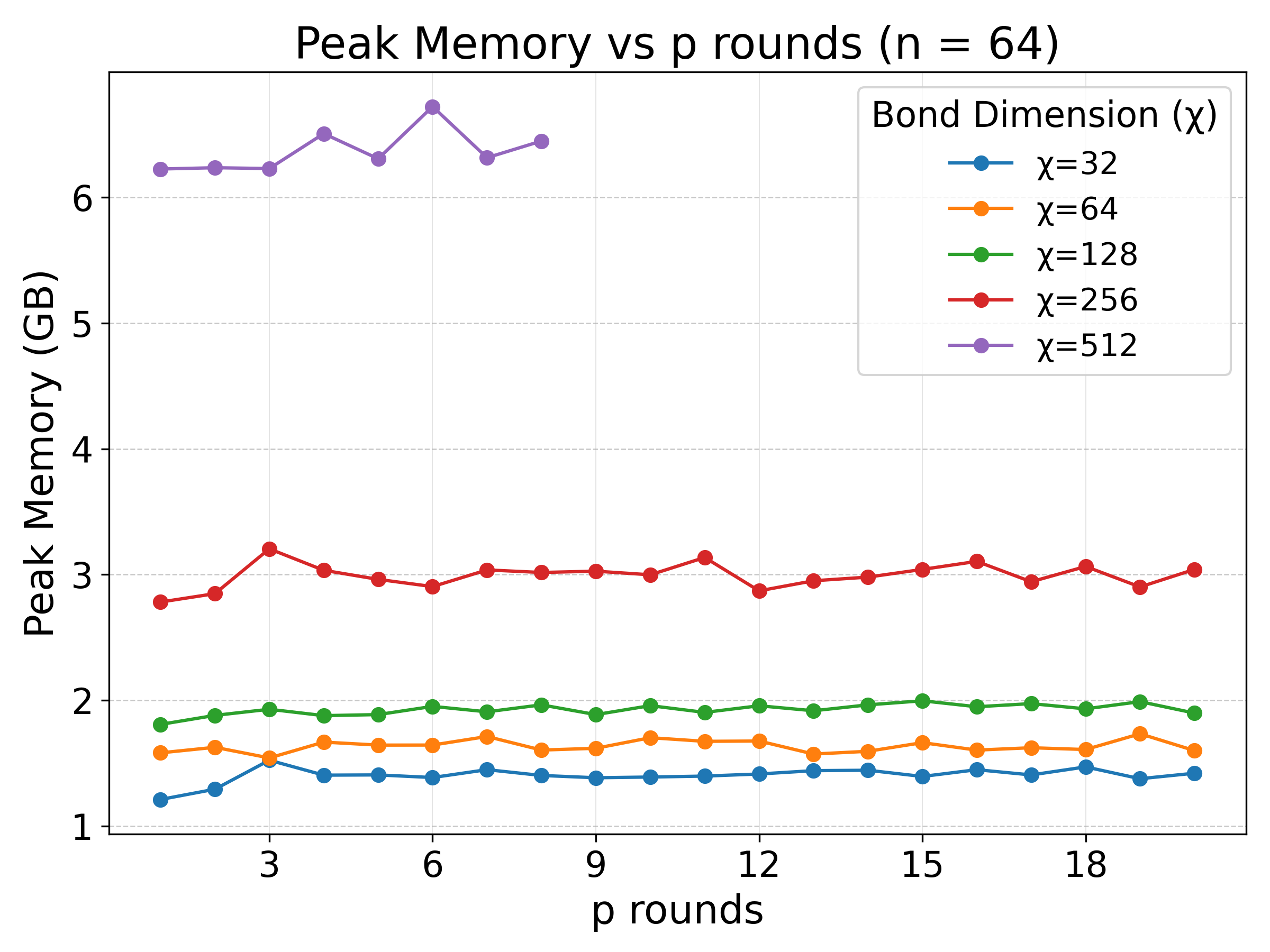} 
  }%
  \hfill
  \subfloat[]{%
    \includegraphics[width=0.49\textwidth]{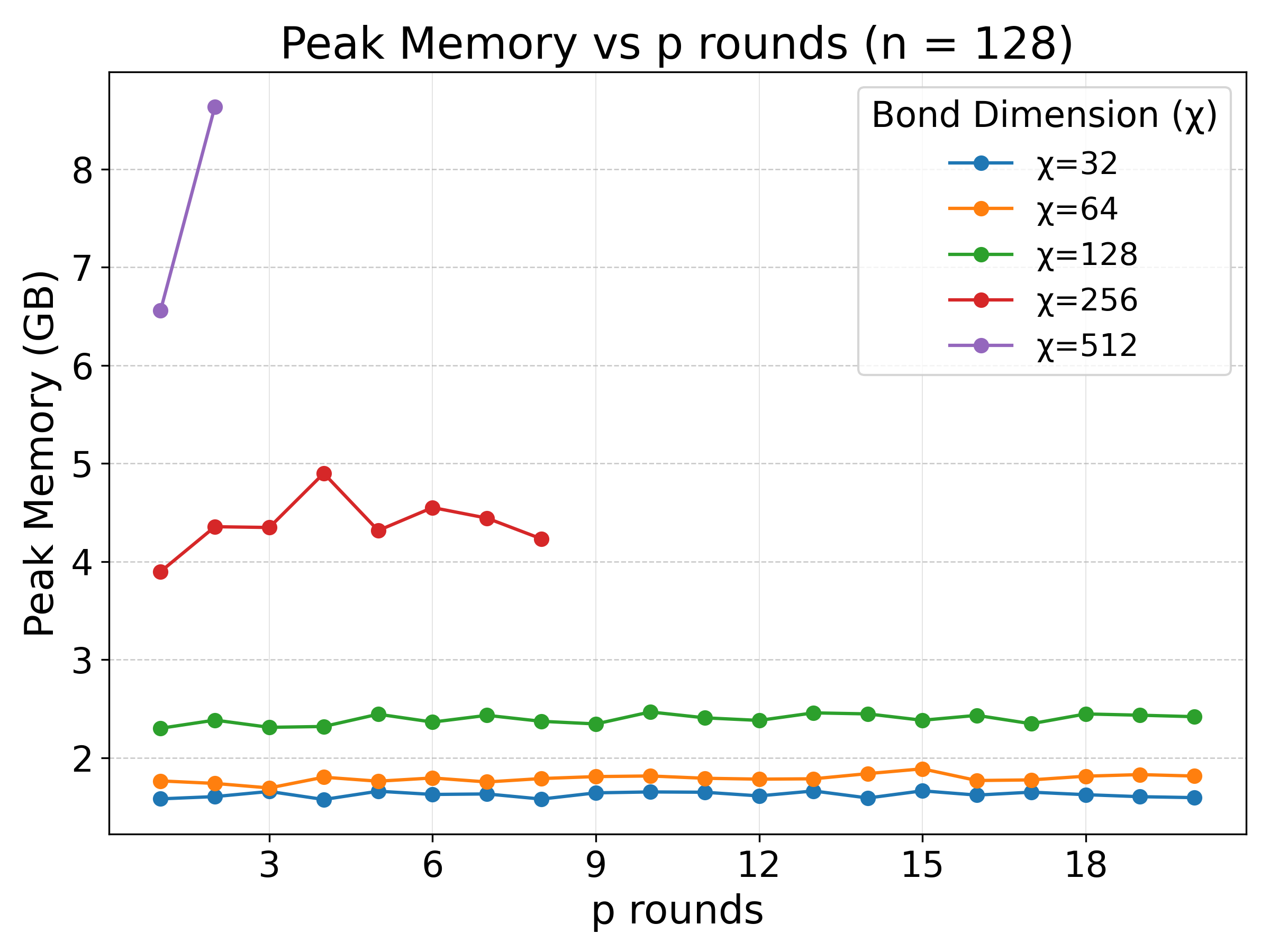}%
  }\\
  \vspace{0.cm} 
  \subfloat[]{%
    \includegraphics[width=0.49\textwidth]{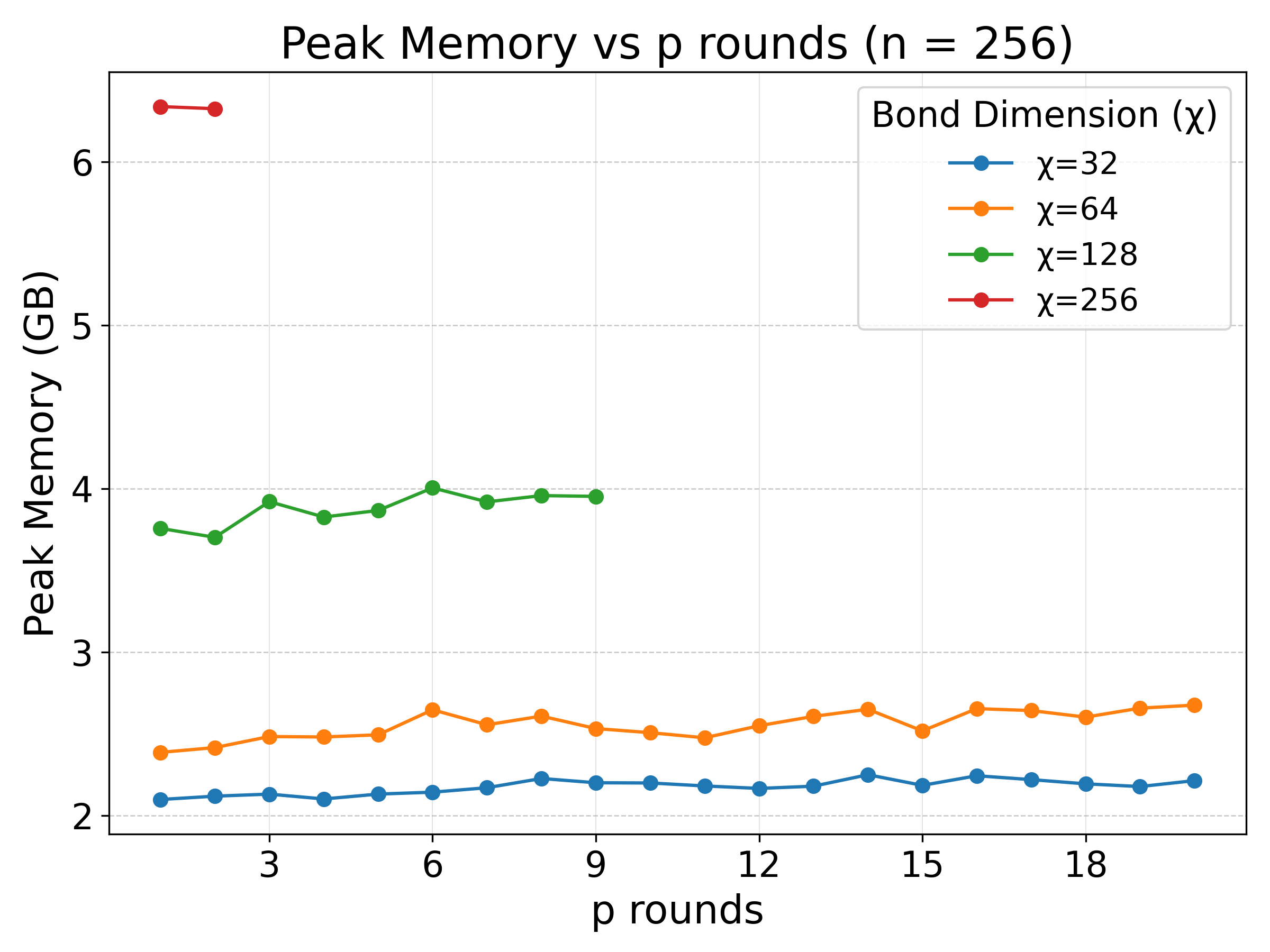}%
  }%
  \hfill
  \subfloat[ ]{%
    \includegraphics[width=0.49\textwidth]{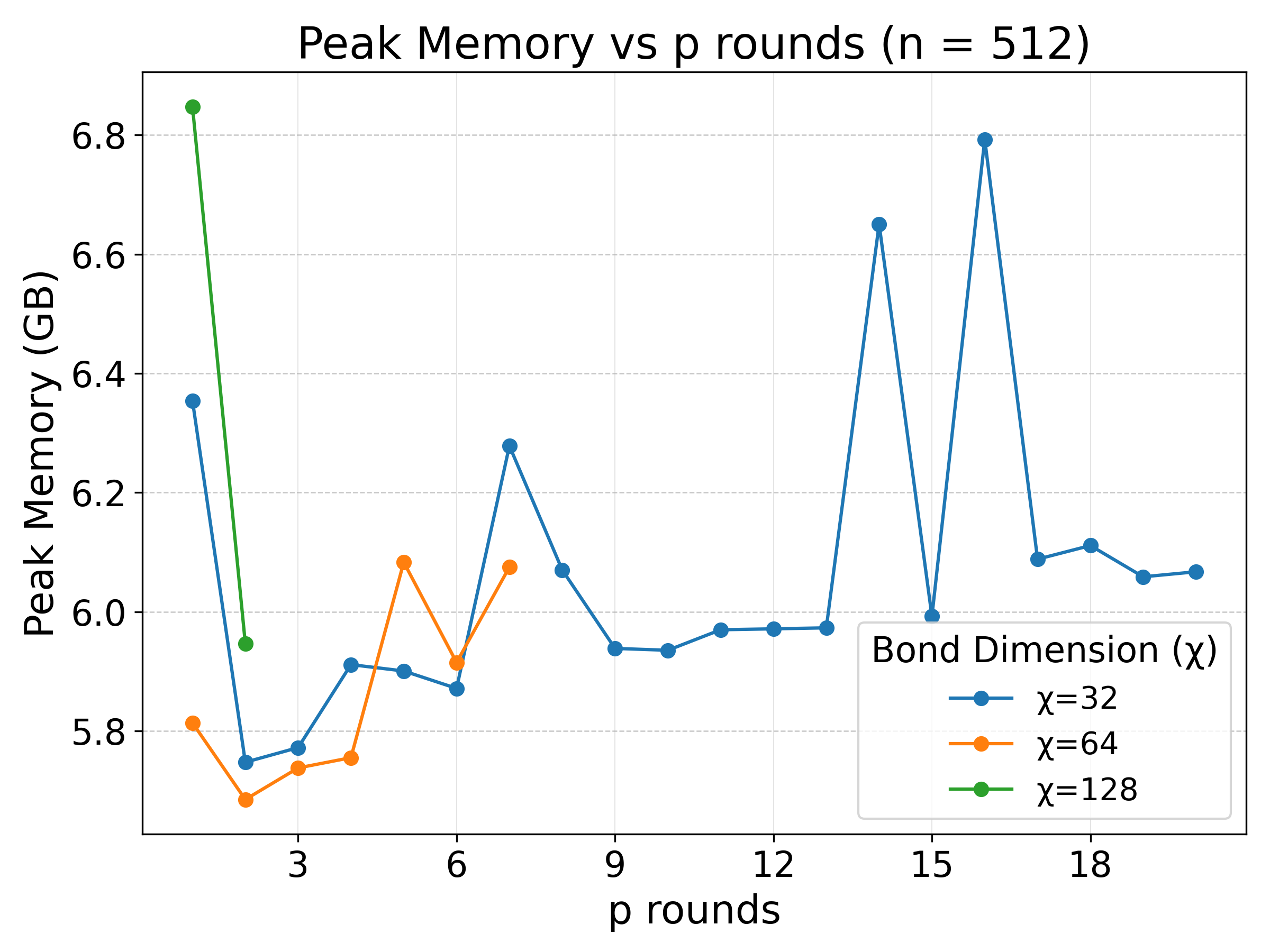}%
  }
  
  \caption{Peak resident memory in Gigabytes (y-axis) vs circuit depth $p$ for bond dimension $\chi \in \{32,64,128,256,512\}$. }
  \label{fig:memory_vs_p}
\end{figure*}

\begin{figure*}[htp]
  \centering
  \subfloat[]{%
    \includegraphics[width=0.49\textwidth]{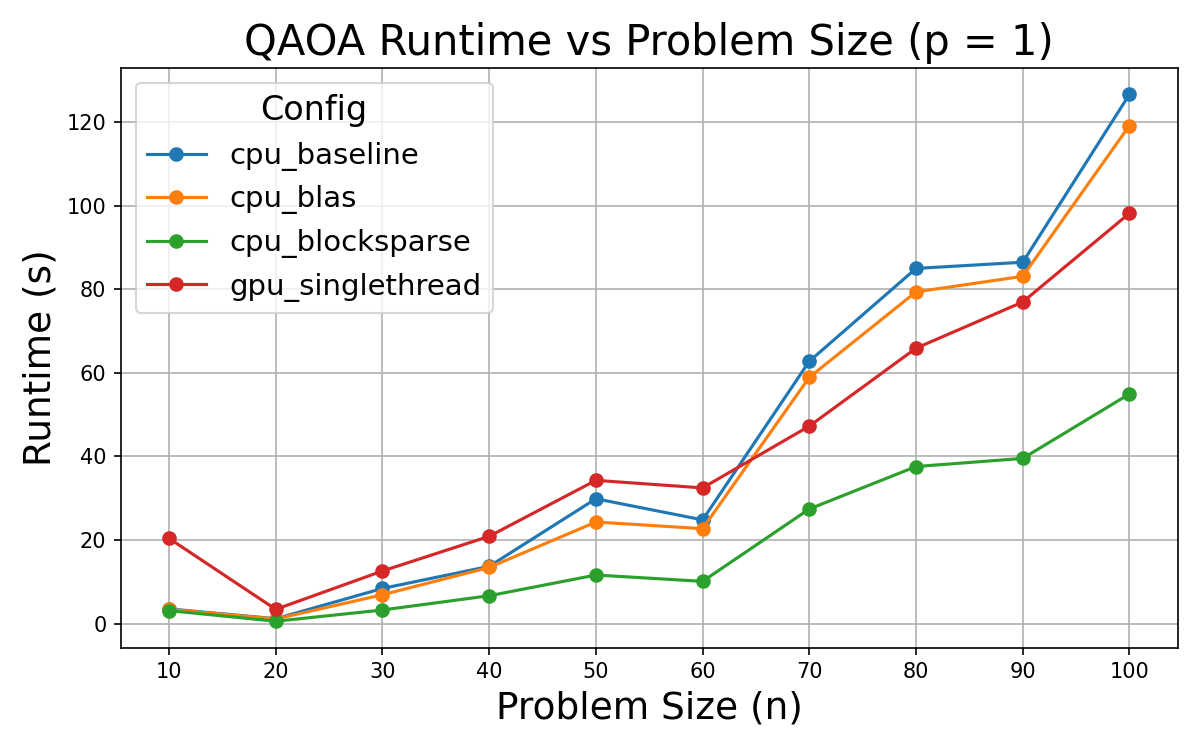} 
  }%
  \hfill
  \subfloat[]{%
    \includegraphics[width=0.49\textwidth]{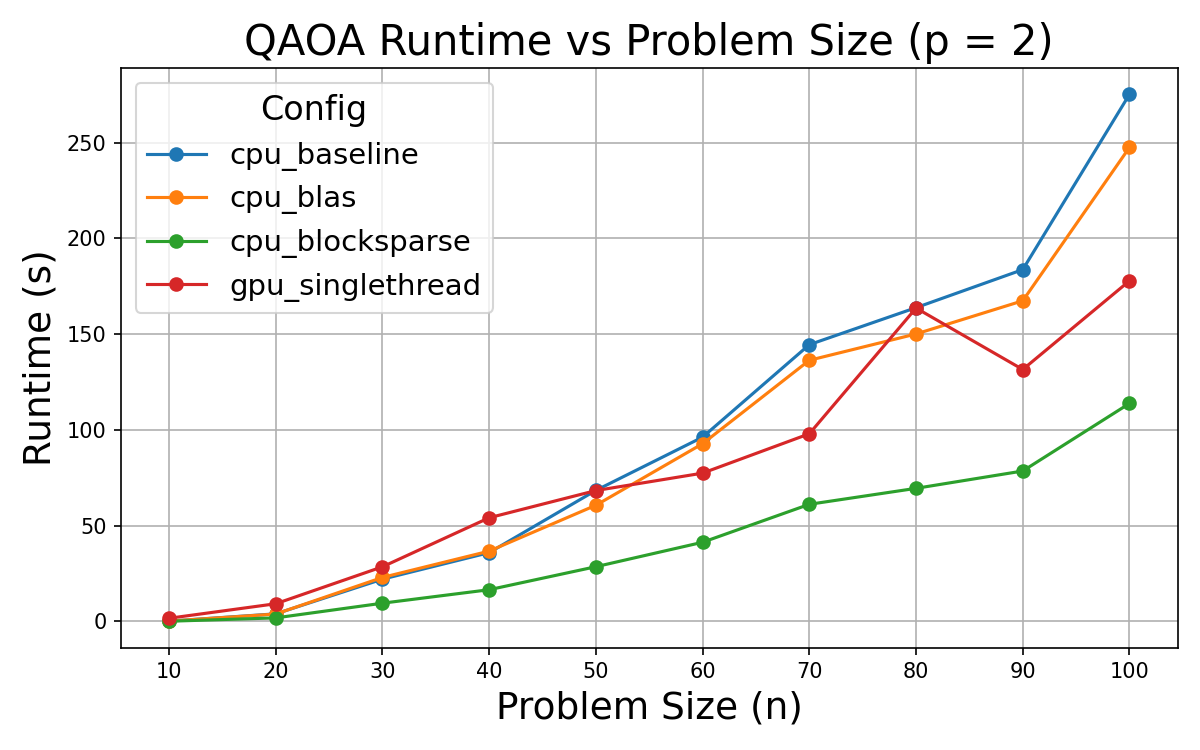}%
  }\\
  \vspace{0.cm} 
  \subfloat[]{%
    \includegraphics[width=0.49\textwidth]{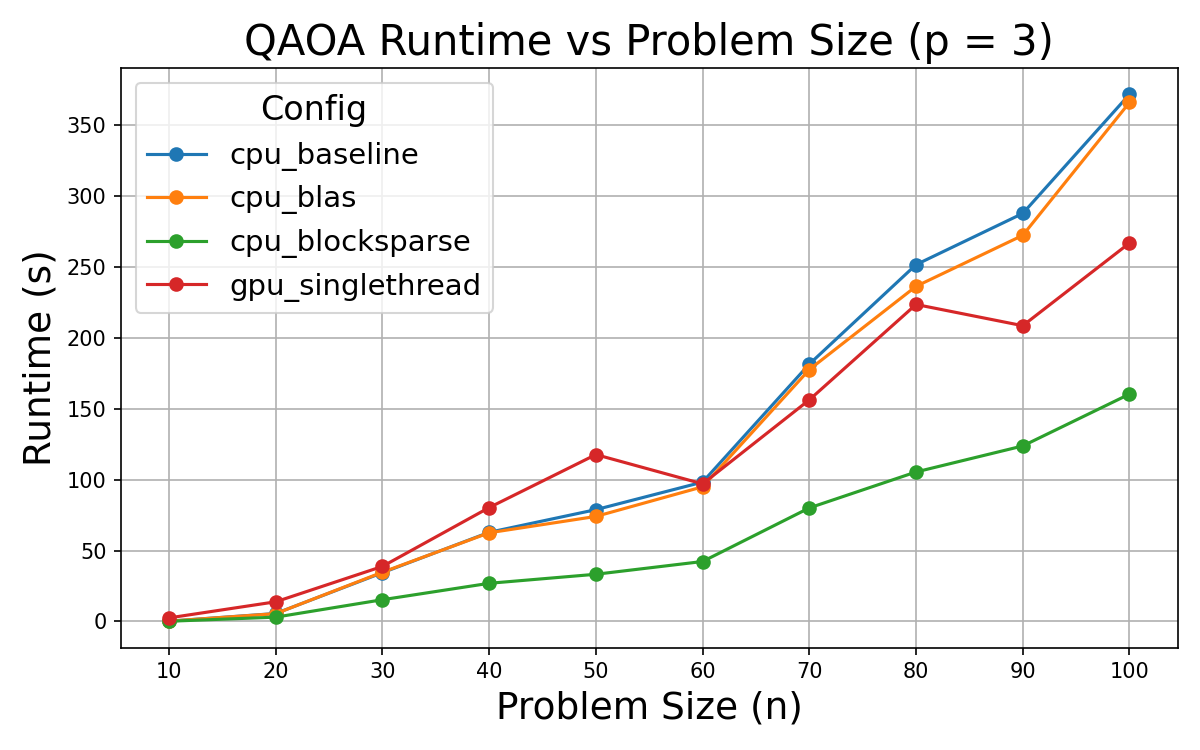}%
  }%
  \hfill
  \subfloat[ ]{%
    \includegraphics[width=0.49\textwidth]{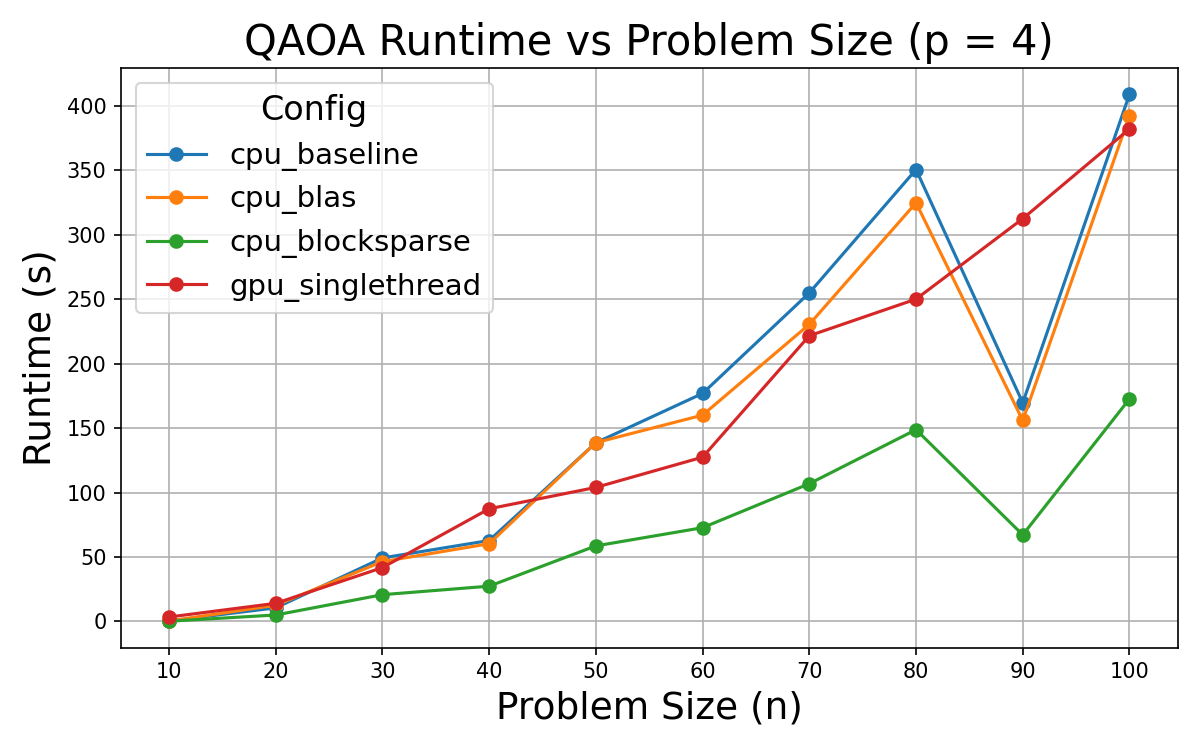}%
  }
  
  \caption{GPU and Multithreading for problem size $n$ (x-axis) vs runtime in seconds (y-axis) for circuit depth $p \in \{1,2,3,4\}$}
  \label{fig:gpu_multi}
\end{figure*}        

\subsection{Runtime vs Bond Dimension}

\label{sec:results_runtime_vs_chi}

 While the previous section concentrated on the effect of circuit depth $p$ on runtime, we now shift focus to the role of bond dimension $\chi$. The bond dimension $\chi$ plays a critical role in the fidelity of MPS simulation by fundamentally governing the number of singular values and thus the entanglement-capturing capability of the MPS representation. To isolate the effect of $\chi$ in Figure \ref{fig:bonddim_vs_runtime}, we fix the QAOA depth $p$ and sweep over a range of bond dimensions for various problem sizes $n$. This orthogonal view enables us to better isolate the computational contributions of bond dimension from those of circuit depth $p$ seen in Figure \ref{fig:runtime_vs_p}.

Figure \ref{fig:bonddim_vs_runtime} presents runtime in seconds (y-axis) as a function of bond dimension $\chi$ (x-axis) for fixed values of circuit depth $p \in \{1, 5, 10, 20\}$. Each subplot corresponds to a fixed $p$, and each curve within a subplot corresponds to a different problem size $n \in \{32, 64, 128, 256, 512\}$.

Across all subplots (a) to (d) in Figure \ref{fig:bonddim_vs_runtime}, we observe what appears to be a superlinear increase in runtime with growing bond dimension $\chi$. This is in line with the theoretical cost of simulating MPS circuits, where the computational complexity of tensor contractions can in worst case scale as $\mathcal{O}(\chi^3)$ \cite{SCHOLLWOCK201196}. For small problem sizes ($n=32$ or $64$), this scaling is more easily sustained across a wider range of $\chi$, while for larger $n$, simulation becomes rapidly infeasible as $\chi$ increases.

For instance, in Figure \ref{fig:bonddim_vs_runtime} (a) where $p = 1$, runtime grows from sub-100 seconds to over 10,000 seconds as $\chi$ increases from $32$ to $128$ for $n=512$. Figures  \ref{fig:bonddim_vs_runtime}(b), \ref{fig:bonddim_vs_runtime}(c), and \ref{fig:bonddim_vs_runtime}(d) reveal that this superlinear trend steepens with circuit depth. At $p = 20$, most curves for large $n$ terminate before reaching $\chi = 256$. While this is not ideal, we do note that the simulations ran on HPC terminated solely due to QOS (Quality of Service) Wall Clock Time constraints, and were not hindered by the memory constraints common in state vector simulation tools.

Problem size $n$ acts as a strong amplifier of runtime. For a fixed $\chi$, increasing $n$ shifts the runtime curve upwards significantly. This is expected, as the number of tensors (and thus the number of operations) grows linearly with $n$, while the complexity of each operation scales polynomially with $\chi$.

Moreover, the effect of increasing $n$ becomes more severe at higher depths, illustrating how the simulation complexity grows along both axes circuit entanglement (via $\chi$) and circuit depth (via $p$). Together, these factors dictate the practical feasibility limits of classical simulation. This reflects a fundamental limitation in classical simulations of quantum circuits, even though the MPS-QAOA tool scales beyond state vector capabilities, larger, deeper, and more entangled systems quickly surpass tractable computational bounds. Nevertheless the MPS-QAOA simulator clearly gives users the ability to scale beyond small system sizes and circuit depths with just the change of a few key word arguments.

While runtime is a primary metric of interest, it is not the main limitation in large-scale quantum circuit simulations that use state vector simulation. Memory consumption is an equally critical metric for scalability, especially as bond dimension grows. Each increase in $\chi$ inflates the size of intermediate and persistent tensors, directly impacting the peak memory footprint. In the next section, we examine how peak resident memory behaves as a function of QAOA depth, helping to clarify the scalability ceiling imposed not only by time but also by available system memory.

\subsection{Memory Trends vs Circuit Depth}

\label{sec:results_memory}

Figure~\ref{fig:memory_vs_p} shows the peak memory usage (in GB) of the MPS-QAOA simulator as a function of circuit depth $p$, for varying problem sizes $n$ and bond dimensions $\chi$. Each subplot fixes a value of $n \in \{64, 128, 256, 512\}$, while each curve within a subplot represents a different bond dimension $\chi$. The peak memory is measured as the maximum resident set size (RSS) during simulation and reflects the highest memory footprint encountered across all time steps.

Unlike the runtime behavior, memory usage remains relatively flat across increasing values of $p$, particularly for smaller problem sizes ($n = 64$ and $n = 128$). For instance, in Figure~\ref{fig:memory_vs_p}(a), the peak memory for each $\chi$ fluctuates slightly with $p$, but shows no clear upward trend. This suggests that for modest system sizes, memory usage is dominated by the static cost of storing MPS tensors and does not grow significantly with QAOA depth.

However, in Figure~\ref{fig:memory_vs_p}(d), which corresponds to $n = 512$, the peak memory usage begins to exhibit more erratic behavior. While the general magnitude stays within the $5.8$ to $6.8$ GB range, the variability indicates that deeper circuits may trigger larger intermediate tensors or auxiliary allocations during contraction, especially in high-entanglement regions.

\subsection{Memory Scaling with Bond Dimension}

Across all subplots, memory consumption increases substantially with bond dimension $\chi$. This is expected since each MPS tensor has dimensions scaling with $\chi$, and storage complexity typically scales as $\mathcal{O}(n \chi^2)$ for 1D chains where $n$ is the number of tensors. This is obvious to see as each MPS site is represented by a 3-dimensional tensor whose scaling is dominated by each of the two bond indices which carry $\chi$ singular values on them i.e. $\chi^{2}$. For example, in Figure~\ref{fig:memory_vs_p}(a), the memory footprint at $\chi = 256$ is roughly $3$ GB, while at $\chi = 512$ it is $6$ GB, a doubling of memory consumption with a doubling of $\chi$.

In subplot ~\ref{fig:memory_vs_p}(d), for $n = 512$, only three values of $\chi$ are shown, and $\chi = 512$ appears only at $p = 1$. This further illustrates the steep memory demands at high bond dimensions. While these memory demands are more exaggerated with larger problem sizes, we note that the lack of data for Figure~\ref{fig:memory_vs_p}(d) was a result of runtime limitations and in fact not memory limitations, though the data is more erratic and less predictable for large system sizes as a result of large tensor contractions.


\subsection{GPU and Multi-threading}
\label{sec:gpu_multithread}
Multi-threading and GPU capabilities are essential for the parallel computing of tensor contractions in any TN-based software. In Figure~\ref{fig:gpu_multi}, the scaling results of the MPS-QAOA simulator are presented. The plots in Figure~\ref{fig:gpu_multi} show problem size $n$ (x-axis) versus runtime (y-axis) for several different implementations of multi-threading, GPU, and single-threaded CPU backends for circuit depth $p \in \{1,2,3,4\}$.

While the \texttt{cpu\_baseline} and \texttt{cpu\_blas} configurations show steady scaling with increasing problem size $n$ and circuit depth $p$, they are consistently outperformed by the \texttt{cpu\_blocksparse} implementation. Across all subfigures ~\ref{fig:gpu_multi}(a)--(d), the \texttt{cpu\_blocksparse} configuration exhibits the lowest runtimes, with particularly pronounced advantages as both $n$ and $p$ increase. Notably, for $p = 4$ and $n = 100$, \texttt{cpu\_blocksparse} completes the task in under 200 seconds, while all other configurations exceed 300 seconds. This demonstrates that the benefits of sparsity-aware multithreading compound significantly at higher tensor contraction complexities.

The superior performance of \texttt{cpu\_blocksparse} stems from its ability to exploit sparsity in the tensor network structure, reducing unnecessary computation and memory overhead. By focusing computational resources only on non-zero elements, it minimizes both arithmetic operations and memory bandwidth usage. Compared to \texttt{cpu\_blas}, which performs dense linear algebra operations regardless of sparsity, \texttt{cpu\_blocksparse} avoids redundant calculations inherent to dense representations. Meanwhile, the \texttt{gpu\_singlethread} configuration underperforms due to underutilization of the GPU’s massive parallelism potential; single-threaded execution on a GPU does not leverage its architectural strengths. Thus, while GPUs are powerful in principle, their efficiency depends heavily on algorithm parallelization and memory access patterns, which \texttt{cpu\_blocksparse} better addresses for this class of TN-contracted quantum circuit simulations.

This highlights a key consideration in building high-performance libraries on top of state-of-the-art toolkits such as \texttt{ITensor}. While \texttt{ITensor}'s multithreaded blocksparse implementation is currently more mature and optimized than its GPU backend, the growing adoption of the \texttt{ITensor} framework suggests that hardware-aware components of \texttt{MPS-JuliQAOA} will increasingly benefit from future improvements in GPU support and other enhanced capabilities. As such, the long-term performance portability and scalability of the simulator are well-aligned with the continued evolution of the underlying tensor library infrastructure.

Overall, these results highlight that for MPS-based QAOA simulations with increasingly large and deep circuits, tailored CPU multithreaded sparse implementations outperform both standard dense CPU libraries and poorly optimized GPU usage. This underscores the importance of both hardware-aware and problem-structure-aware software design in high-performance quantum optimization circuit simulation.

\section{Conclusions and Future Work}
\label{sec:conclusion}

In this work, we presented \texttt{MPS-JuliQAOA}, a high-performance simulation framework for QAOA built in Julia using the \texttt{ITensor} ecosystem. Our implementation supports arbitrary diagonal cost Hamiltonians, leverages differentiable programming via \texttt{Zygote.jl} for gradient-based optimization, and enables scalable circuit simulation through Matrix Product State (MPS) techniques. We demonstrated a modular architecture that seamlessly integrates Hamiltonian specification, tensor network simulation, and parameter optimization, making the software flexible, extensible, and user-friendly for a wide variety of combinatorial optimization problems.

Through an extensive benchmarking suite, we evaluated runtime and memory behavior across problem sizes, bond dimensions, and circuit depths. The results validate the linear scaling of runtime with QAOA depth $p$ for fixed bond dimension $\chi$, while illustrating the superlinear scaling with $\chi$ itself. Memory usage followed expected trends of $\mathcal{O}(n \chi^2)$, with larger $\chi$ being the primary limitation in simulation feasibility. Notably, we showed that ITensor's multithreaded \texttt{cpu\_blocksparse} backend consistently outperforms both BLAS-accelerated and GPU single-threaded alternatives for tensor contraction-heavy workloads, highlighting the importance of exploiting sparsity in tensor structures.

These findings underscore the critical role of hardware-aware optimizations in classical simulation of variational quantum algorithms. As ITensor’s GPU support matures, and with continued advancements in parallel tensor algebra, the \texttt{MPS-JuliQAOA} simulator stands to benefit further extending its applicability to even larger and more entangled systems. This work sets the stage for future explorations in hybrid classical-quantum workflows, optimized QAOA compilation, and problem-specific Hamiltonian engineering using scalable tensor network methods. It also helps to fill the gap in classical quantum software simulation tools for QAOA optimization tasks

There are plans to support continued development of MPS-JuliQAOA, both in terms of ease-of-use (e.g. option to maximize objectives without needing to negate Hamiltonian coefficients, additional angle-finding settings, additional helper functions for other optimization problems, etc), and increased functionality (e.g. implementing alternative mixers, allowing non-diagonal cost Hamiltonians, exploiting light-cone arguments to reduce runtime and memory resources, etc).

Additionally, much of the current QAOA literature includes statevector simulation results that make claims regarding scaling as the number of qubits increases. However, it is difficult to be convinced of such claims due to the limited qubit amounts that are achievable via statevector simulation; on the other hand, our tool can be used to quickly bolster such claims by showing that the empirically-found trends continue to hold for larger system sizes.

\section{Acknowledgments}
\label{section:acknowledgments}
This work was supported by the U.S. Department of Energy through the Los Alamos National Laboratory. Los Alamos National Laboratory is operated by Triad National Security, LLC, for the National Nuclear Security Administration of U.S. Department of Energy (Contract No. 89233218CNA000001). The research presented in this article was supported by the Laboratory Directed Research and Development program of Los Alamos National Laboratory under project number 20230049DR as well as by the NNSA's Advanced Simulation and Computing Beyond Moore's Law Program at Los Alamos National Laboratory. This work has been assigned LANL technical report number LA-UR-25-27542.

\vspace{12pt}

\appendices

\section{MIS Interactions Derivation}
\label{sec:MIS_appendix}
Recall from Equation \ref{eqn:MIS_Ham} that we can write the Maximum Independent Set (MIS) Hamiltonian (with penalty term $\lambda$) as $H_C = H_1 + H_2$ where,

$$H_1 = \frac{1}{2}\sum_{i \in V} (I - Z_i),$$
$$H_2 = - \frac{\lambda}{4} \sum_{(i,j) \in E} (I - Z_i)(I - Z_j).$$

Expanding $H_1$, we have
$$H_1 = \frac{n}{2}I + \sum_{i \in V} \frac{-1}{2}Z_i.$$

Before expanding $H_2$, we first make an important observation:
$$\sum_{(i,j) \in E} (Z_i + Z_j) = \sum_{i \in V} \text{deg}_i Z_i,$$
where $\text{deg}_i$ is the degree of vertex $i$ in the graph; the proof of this observation can easily be seen as an adaptation of the proof of the well-known handshaking lemma.

Using the observation above, we now expand $H_2$:
\begin{align*}
    H_2 &=  - \frac{\lambda}{4} \sum_{(i,j) \in E} (I - Z_i)(I - Z_j)\\
    &= \frac{-\lambda}{4} \sum_{(i,j) \in E} (I - (Z_i + Z_j) + Z_iZ_j)\\
    &= \frac{-\lambda m}{4} I  + \frac{\lambda}{4}\sum_{(i,j) \in E} (Z_i + Z_j) + \sum_{(i,j)\in E} \frac{-\lambda}{4}Z_iZ_j\\
    &=  \frac{-\lambda m}{4} I  + \frac{\lambda}{4}\sum_{i \in V} \text{deg}_i Z_i + \sum_{(i,j)\in E} \frac{-\lambda}{4}Z_iZ_j.
\end{align*}

Combining the expansions of $H_1$ and $H_2$ then yields the MIS Hamiltonian in the desired form:
$$H_C = kI + \sum_{i \in V} k_i Z_i + \sum_{(i,j) \in E} k_{ij} Z_iZ_j,$$
where,
$$k = \frac{n}{2} - \frac{\lambda}{4}m$$
$$k_i = \frac{\lambda  }{4}\text{deg}_i - \frac{1}{2}$$
$$k_{ij} = -\frac{\lambda}{4}.$$

\bibliographystyle{IEEEtran}
\bibliography{sources}

\begin{thebibliography}{10}
\providecommand{\url}[1]{#1}
\csname url@samestyle\endcsname
\providecommand{\newblock}{\relax}
\providecommand{\bibinfo}[2]{#2}
\providecommand{\BIBentrySTDinterwordspacing}{\spaceskip=0pt\relax}
\providecommand{\BIBentryALTinterwordstretchfactor}{4}
\providecommand{\BIBentryALTinterwordspacing}{\spaceskip=\fontdimen2\font plus
\BIBentryALTinterwordstretchfactor\fontdimen3\font minus
  \fontdimen4\font\relax}
\providecommand{\BIBforeignlanguage}[2]{{%
\expandafter\ifx\csname l@#1\endcsname\relax
\typeout{** WARNING: IEEEtran.bst: No hyphenation pattern has been}%
\typeout{** loaded for the language `#1'. Using the pattern for}%
\typeout{** the default language instead.}%
\else
\language=\csname l@#1\endcsname
\fi
#2}}
\providecommand{\BIBdecl}{\relax}
\BIBdecl

\bibitem{di2024quantum}
A.~Di~Meglio, K.~Jansen, I.~Tavernelli, C.~Alexandrou, S.~Arunachalam, C.~W.
  Bauer, K.~Borras, S.~Carrazza, A.~Crippa, V.~Croft \emph{et~al.}, ``Quantum
  computing for high-energy physics: State of the art and challenges,''
  \emph{PRX Quantum}, vol.~5, no.~3, p. 037001, 2024.

\bibitem{cao2019quantum}
Y.~Cao, J.~Romero, J.~P. Olson, M.~Degroote, P.~D. Johnson, M.~Kieferov{\'a},
  I.~D. Kivlichan, T.~Menke, B.~Peropadre, N.~P. Sawaya \emph{et~al.},
  ``Quantum chemistry in the age of quantum computing,'' \emph{Chemical
  reviews}, vol. 119, no.~19, pp. 10\,856--10\,915, 2019.

\bibitem{xu2023herculeantaskclassicalsimulation}
\BIBentryALTinterwordspacing
X.~Xu, S.~Benjamin, J.~Sun, X.~Yuan, and P.~Zhang, ``A herculean task:
  Classical simulation of quantum computers,'' 2023. [Online]. Available:
  \url{https://arxiv.org/abs/2302.08880}
\BIBentrySTDinterwordspacing

\bibitem{365700}
P.~Shor, ``Algorithms for quantum computation: discrete logarithms and
  factoring,'' in \emph{Proceedings 35th Annual Symposium on Foundations of
  Computer Science}, 1994, pp. 124--134.

\bibitem{10.1145/237814.237866}
\BIBentryALTinterwordspacing
L.~K. Grover, ``A fast quantum mechanical algorithm for database search,'' in
  \emph{Proceedings of the Twenty-Eighth Annual ACM Symposium on Theory of
  Computing}, ser. STOC '96.\hskip 1em plus 0.5em minus 0.4em\relax New York,
  NY, USA: Association for Computing Machinery, 1996, p. 212–219. [Online].
  Available: \url{https://doi.org/10.1145/237814.237866}
\BIBentrySTDinterwordspacing

\bibitem{Peruzzo_2014}
\BIBentryALTinterwordspacing
A.~Peruzzo, J.~McClean, P.~Shadbolt, M.-H. Yung, X.-Q. Zhou, P.~J. Love,
  A.~Aspuru-Guzik, and J.~L. O’Brien, ``A variational eigenvalue solver on a
  photonic quantum processor,'' \emph{Nature Communications}, vol.~5, no.~1,
  Jul. 2014. [Online]. Available: \url{http://dx.doi.org/10.1038/ncomms5213}
\BIBentrySTDinterwordspacing

\bibitem{farhi2014quantumapproximateoptimizationalgorithm}
\BIBentryALTinterwordspacing
E.~Farhi, J.~Goldstone, and S.~Gutmann, ``A quantum approximate optimization
  algorithm,'' 2014. [Online]. Available: \url{https://arxiv.org/abs/1411.4028}
\BIBentrySTDinterwordspacing

\bibitem{cerezo2021variational}
M.~Cerezo, A.~Arrasmith, R.~Babbush, S.~C. Benjamin, S.~Endo, K.~Fujii, J.~R.
  McClean, K.~Mitarai, X.~Yuan, L.~Cincio \emph{et~al.}, ``Variational quantum
  algorithms,'' \emph{Nature Reviews Physics}, vol.~3, no.~9, pp. 625--644,
  2021.

\bibitem{berezutskii2025tensor}
A.~Berezutskii, M.~Liu, A.~Acharya, R.~Ellerbrock, J.~Gray, R.~Haghshenas,
  Z.~He, A.~Khan, V.~Kuzmin, D.~Lyakh \emph{et~al.}, ``Tensor networks for
  quantum computing,'' \emph{arXiv preprint arXiv:2503.08626}, 2025.

\bibitem{verstraete2004renormalization}
F.~Verstraete and J.~I. Cirac, ``Renormalization algorithms for quantum-many
  body systems in two and higher dimensions,'' \emph{arXiv preprint
  cond-mat/0407066}, 2004.

\bibitem{cincio2008multiscale}
L.~Cincio, J.~Dziarmaga, and M.~M. Rams, ``Multiscale entanglement
  renormalization ansatz in two dimensions: quantum ising model,''
  \emph{Physical review letters}, vol. 100, no.~24, p. 240603, 2008.

\bibitem{PhysRevLett.69.2863}
\BIBentryALTinterwordspacing
S.~R. White, ``Density matrix formulation for quantum renormalization groups,''
  \emph{Phys. Rev. Lett.}, vol.~69, pp. 2863--2866, Nov 1992. [Online].
  Available: \url{https://link.aps.org/doi/10.1103/PhysRevLett.69.2863}
\BIBentrySTDinterwordspacing

\bibitem{PAECKEL2019167998}
\BIBentryALTinterwordspacing
S.~Paeckel, T.~Köhler, A.~Swoboda, S.~R. Manmana, U.~Schollwöck, and
  C.~Hubig, ``Time-evolution methods for matrix-product states,'' \emph{Annals
  of Physics}, vol. 411, p. 167998, 2019. [Online]. Available:
  \url{https://www.sciencedirect.com/science/article/pii/S0003491619302532}
\BIBentrySTDinterwordspacing

\bibitem{Golden_2023}
\BIBentryALTinterwordspacing
J.~Golden, A.~Baertschi, D.~O’Malley, E.~Pelofske, and S.~Eidenbenz,
  ``Juliqaoa: Fast, flexible qaoa simulation,'' in \emph{Proceedings of the SC
  ’23 Workshops of the International Conference on High Performance
  Computing, Network, Storage, and Analysis}, ser. SC-W 2023.\hskip 1em plus
  0.5em minus 0.4em\relax ACM, Nov. 2023, p. 1454–1459. [Online]. Available:
  \url{http://dx.doi.org/10.1145/3624062.3624220}
\BIBentrySTDinterwordspacing

\bibitem{qiskit2024}
A.~Javadi-Abhari, M.~Treinish, K.~Krsulich, C.~J. Wood, J.~Lishman, J.~Gacon,
  S.~Martiel, P.~D. Nation, L.~S. Bishop, A.~W. Cross, B.~R. Johnson, and J.~M.
  Gambetta, ``Quantum computing with {Q}iskit,'' 2024.

\bibitem{Bode2023}
\BIBentryALTinterwordspacing
T.~Bode, D.~Bagrets, A.~Misra-Spieldenner, T.~Stollenwerk, and F.~K. Wilhelm,
  ``Qaoa.jl: Toolkit for the quantum and mean-field approximate optimization
  algorithms,'' \emph{Journal of Open Source Software}, vol.~8, no.~86, p.
  5364, 2023. [Online]. Available: \url{https://doi.org/10.21105/joss.05364}
\BIBentrySTDinterwordspacing

\bibitem{PRXQuantum.4.030335}
\BIBentryALTinterwordspacing
A.~Misra-Spieldenner, T.~Bode, P.~K. Schuhmacher, T.~Stollenwerk, D.~Bagrets,
  and F.~K. Wilhelm, ``Mean-field approximate optimization algorithm,''
  \emph{PRX Quantum}, vol.~4, p. 030335, Sep 2023. [Online]. Available:
  \url{https://link.aps.org/doi/10.1103/PRXQuantum.4.030335}
\BIBentrySTDinterwordspacing

\bibitem{ITensor}
\BIBentryALTinterwordspacing
M.~Fishman, S.~R. White, and E.~M. Stoudenmire, ``{The ITensor Software Library
  for Tensor Network Calculations},'' \emph{SciPost Phys. Codebases}, p.~4,
  2022. [Online]. Available:
  \url{https://scipost.org/10.21468/SciPostPhysCodeb.4}
\BIBentrySTDinterwordspacing

\bibitem{Lykov_2021}
\BIBentryALTinterwordspacing
D.~Lykov, A.~Chen, H.~Chen, K.~Keipert, Z.~Zhang, T.~Gibbs, and Y.~Alexeev,
  ``Performance evaluation and acceleration of the qtensor quantum circuit
  simulator on gpus,'' in \emph{2021 IEEE/ACM Second International Workshop on
  Quantum Computing Software (QCS)}.\hskip 1em plus 0.5em minus 0.4em\relax
  IEEE, Nov. 2021, p. 27–34. [Online]. Available:
  \url{http://dx.doi.org/10.1109/QCS54837.2021.00007}
\BIBentrySTDinterwordspacing

\bibitem{Bridgeman_2017}
\BIBentryALTinterwordspacing
J.~C. Bridgeman and C.~T. Chubb, ``Hand-waving and interpretive dance: an
  introductory course on tensor networks,'' \emph{Journal of Physics A:
  Mathematical and Theoretical}, vol.~50, no.~22, p. 223001, May 2017.
  [Online]. Available: \url{http://dx.doi.org/10.1088/1751-8121/aa6dc3}
\BIBentrySTDinterwordspacing

\bibitem{biamonte2017tensornetworksnutshell}
\BIBentryALTinterwordspacing
J.~Biamonte and V.~Bergholm, ``Tensor networks in a nutshell,'' 2017. [Online].
  Available: \url{https://arxiv.org/abs/1708.00006}
\BIBentrySTDinterwordspacing

\bibitem{Or_s_2014}
\BIBentryALTinterwordspacing
R.~Orús, ``A practical introduction to tensor networks: Matrix product states
  and projected entangled pair states,'' \emph{Annals of Physics}, vol. 349, p.
  117–158, Oct. 2014. [Online]. Available:
  \url{http://dx.doi.org/10.1016/j.aop.2014.06.013}
\BIBentrySTDinterwordspacing

\bibitem{Verstraete_2006}
\BIBentryALTinterwordspacing
F.~Verstraete and J.~I. Cirac, ``Matrix product states represent ground states
  faithfully,'' \emph{Physical Review B}, vol.~73, no.~9, Mar. 2006. [Online].
  Available: \url{http://dx.doi.org/10.1103/PhysRevB.73.094423}
\BIBentrySTDinterwordspacing

\bibitem{feeney2024bettersolutionprobabilitymetric}
\BIBentryALTinterwordspacing
S.~Feeney, R.~Tate, and S.~Eidenbenz, ``The better solution probability metric:
  Optimizing qaoa to outperform its warm-start solution,'' 2024. [Online].
  Available: \url{https://arxiv.org/abs/2409.09012}
\BIBentrySTDinterwordspacing

\bibitem{kus2024gradientfreeneuraltopologyoptimization}
\BIBentryALTinterwordspacing
G.~Kus and M.~A. Bessa, ``Gradient-free neural topology optimization: Towards
  effective fracture-resistant designs,'' 2024. [Online]. Available:
  \url{https://arxiv.org/abs/2403.04937}
\BIBentrySTDinterwordspacing

\bibitem{baydin2018automaticdifferentiationmachinelearning}
\BIBentryALTinterwordspacing
A.~G. Baydin, B.~A. Pearlmutter, A.~A. Radul, and J.~M. Siskind, ``Automatic
  differentiation in machine learning: a survey,'' 2018. [Online]. Available:
  \url{https://arxiv.org/abs/1502.05767}
\BIBentrySTDinterwordspacing

\bibitem{fang2024stepbystepintroductionimplementationautomatic}
\BIBentryALTinterwordspacing
Y.-H. Fang, H.-Z. Lin, J.-J. Liu, and C.-J. Lin, ``A step-by-step introduction
  to the implementation of automatic differentiation,'' 2024. [Online].
  Available: \url{https://arxiv.org/abs/2402.16020}
\BIBentrySTDinterwordspacing

\bibitem{Guo_2023}
\BIBentryALTinterwordspacing
C.~Guo, Y.~Fan, Z.~Xu, and H.~Shang, ``Differentiable matrix product states for
  simulating variational quantum computational chemistry,'' \emph{Quantum},
  vol.~7, p. 1192, Dec. 2023. [Online]. Available:
  \url{http://dx.doi.org/10.22331/q-2023-12-04-1192}
\BIBentrySTDinterwordspacing

\bibitem{hadfield2021representation}
S.~Hadfield, ``On the representation of boolean and real functions as
  hamiltonians for quantum computing,'' \emph{ACM Transactions on Quantum
  Computing}, vol.~2, no.~4, pp. 1--21, 2021.

\bibitem{lucas2014ising}
A.~Lucas, ``Ising formulations of many np problems,'' \emph{Frontiers in
  physics}, vol.~2, p.~5, 2014.

\bibitem{SCHOLLWOCK201196}
\BIBentryALTinterwordspacing
U.~Schollwöck, ``The density-matrix renormalization group in the age of matrix
  product states,'' \emph{Annals of Physics}, vol. 326, no.~1, pp. 96--192,
  2011, january 2011 Special Issue. [Online]. Available:
  \url{https://www.sciencedirect.com/science/article/pii/S0003491610001752}
\BIBentrySTDinterwordspacing

\end{thebibliography}
\end{document}